\documentclass[aps,prb,twocolumn,floats,epsfig,superscriptaddress]{revtex4}
\usepackage[iso-8859-1]{inputenx}
\usepackage{amssymb}
\usepackage{amsbsy}
\usepackage{amsmath}
\usepackage{xcolor}
\usepackage{epsfig}
\usepackage{color}
\usepackage{soul}
\usepackage{float}
\usepackage[colorlinks,linktocpage,bookmarks=false,citecolor=blue,linkcolor=red,urlcolor=blue]{hyperref}

\newcommand{\ing}{\includegraphics}
\newcommand{\bib}{\bibitem}
\newcommand{\beq}{\begin{equation}}
\newcommand{\eeq}{\end{equation}}
\newcommand{\bea}{\begin{eqnarray}}
\newcommand{\eea}{\end{eqnarray}}

\newcommand{\bra}[1]{\langle#1|}
\newcommand{\ket}[1]{|#1\rangle}

\usepackage{xcolor}

\begin{document}

\title{Signatures of multifractality in a periodically driven interacting Aubry-Andr\'{e} model}

\author{Madhumita Sarkar}

\affiliation{School of Physical Sciences, Indian
Association for the Cultivation of Science, 2A and 2B Raja S. C.
Mullick Road, Jadavpur 700032, India}
\author{ Roopayan Ghosh}
\affiliation{School of Physical Sciences, Indian
Association for the Cultivation of Science, 2A and 2B Raja S. C.
Mullick Road, Jadavpur 700032, India}
\affiliation{Department of Physics, FMF, University of Ljubljana, Jadranska 19, SI-1000 Ljubljana, Slovenia}
\author{Arnab Sen}
\affiliation{School of Physical Sciences, Indian
Association for the Cultivation of Science, 2A and 2B Raja S. C.
Mullick Road, Jadavpur 700032, India}
\author{K. Sengupta}
\affiliation{School of Physical Sciences, Indian
Association for the Cultivation of Science, 2A and 2B Raja S. C.
Mullick Road, Jadavpur 700032, India}
\date{\today}

\date{\today}

\begin{abstract}

We study the many-body localization (MBL) transition of Floquet
eigenstates in a driven, interacting fermionic chain with an
incommensurate Aubry-Andr\'{e} potential and a time-periodic hopping
amplitude as a function of the drive frequency $\omega_D$ using
exact diagonalization (ED). We find that the nature of the Floquet
eigenstates change from ergodic to Floquet-MBL with increasing
frequency; moreover, for a significant range of intermediate
$\omega_D$, the Floquet eigenstates exhibit non-trivial fractal
dimensions. We find a possible transition from the ergodic to this
multifractal phase followed by a gradual crossover to the MBL phase
as the drive frequency is increased. We also study the fermion
auto-correlation function, entanglement entropy, normalized
participation ratio (NPR), fermion transport and the inverse
participation ratio (IPR) as a function of $\omega_D$. We show that
the auto-correlation, fermion transport and NPR displays
qualitatively different characteristics (compared to their behavior
in the ergodic and MBL regions) for the range of $\omega_D$ which
supports multifractal eigenstates. In contrast, the entanglement
growth in this frequency range tend to have similar features as in
the MBL regime; its rate of growth is controlled by $\omega_D$. Our
analysis thus indicates that the multifractal nature of Floquet-MBL
eigenstates can be detected by studying auto-correlation function
and fermionic transport of these driven chains. We support our
numerical results with a semi-analytic expression of the Floquet
Hamiltonian obtained using Floquet perturbation theory (FPT) and
discuss possible experiments which can test our predictions.

\end{abstract}

\maketitle

\section{Introduction}
It is well-known that non-interacting fermions in one-dimension with
short range hopping exhibits localization for arbitrary weak
disorder potential \cite{rev1a,rev12}. In contrast, fermion chains
subjected to quasiperiodic potential exhibit a
localization-delocalization transition at a finite potential
strength
\cite{aaref,aaref2,aaref3,aaref4,gaaref,gaaref2,gaaref3,gaaref4,gaaref5,gaaref6,gaaref7,gaaref8,gaaref9}.
Localization in such 1D fermion chains with quasiperiodic potentials
has been studied extensively in the past
\cite{aaref,aaref2,aaref3,aaref4,
gaaref,gaaref2,gaaref3,gaaref4,gaaref5,gaaref6,gaaref7,gaaref8,gaaref9,sinharef,
otherref,Zu,otherref2}. In recent times, such systems have also been
experimentally realized using ultracold atom chains
\cite{exp1,exp12,exp2,exp22}. The simplest of such models with
quasiperiodic potential is termed as Aubry-Andr\'{e} (AA) model
\cite{aaref,aaref2,aaref3,aaref4}. The Hamiltonian the AA model is
given by
\begin{eqnarray}
H_{\rm NI} &=& H_0 + H_A \nonumber\\
H_0 &=& - \frac{\mathcal{J}}{2}\sum_j c_j^{\dagger} (c_{j+1} +
c_{j-1}) \nonumber\\
H_A &=& \sum_j V_0 \cos(2\pi \eta j + \phi) c_j^{\dagger} c_j
\end{eqnarray}
where $c_j$ denotes fermionic annihilation operator at site $j$,
$\hat n_j=c_j^{\dagger} c_j$ is the corresponding fermion number
operator, $\mathcal{J}$ is the nearest-neighbor hopping amplitude of
the fermions, $\eta$ is an irrational number usually chosen to be
the golden ratio $(\sqrt{5}-1)/2$, $V_0$ is the amplitude of the AA
potential, and $\phi$ is an arbitrary global phase. The model
exhibits a localization-delocalization transition at $V_0={\mathcal
J}$.

More recently, non-equilibrium dynamics of interacting quantum
systems has been extensively studied
\cite{rev1,rev2,rev3,rev4,rev5,rev6,rev7, rev8}. A class of such
studies has concentrated on periodic drive for which the properties
of the system is controlled by its Floquet Hamiltonian \cite{rev9}.
The Floquet Hamiltonian $H_F$ of a periodically driven system
contains information about its properties at stroboscopic times $n
T$, where $T= 2 \pi/\omega_D$ is the drive period, $\omega_D$ is the
drive frequency, and $n$ is an integer. This feature stems from the
fact the evolution operator for such systems satisfy $U(nT,0)=
\exp[-i n H_F T/\hbar]$, where $\hbar$ is the Planck's constant. It
is well known that an interacting quantum systems without the
presence of quasiperiodic potential or disorder also undergoes
dynamical localization\cite{dynloc1a, dynloc1b, dynloc1c,dynloc1d,
dynloc2a, dynloc2b, dynloc3}, exhibits dynamical freezing
\cite{dynf1a,dynf1b,dynf1c,dynf2a,dynf2b,dynf3}, and can display
violation of eigenstate thermalization hypothesis (ETH)
\cite{eth1,eth2,eth3,eth4} due to quantum scars
\cite{ethv0,ethv01,ethv02,ethv03,
ethv04,ethv05,ethv06,ethv07,ethv08,ethv09} whose signature can also
be found using periodic drives \cite{ethv1,ethv2}. However, the
origin of such drive controlled localization or ETH violation is
quite different from that found in traditional many-body
localization (MBL)
\cite{mbl1,mbl11,mbl12,mbl13,mbl14,mbl15,mbl16,mbl17,mbl18,mbl19,mbl20,
mbl21,mbl22,mbl23,mbl24,mbl25,mbl26,mbl27,mbl28,mbl29,mbl30,mbl31,
rafael,abanin,pal,huse,znidaric1,rigol,chalker,lin,gopalakrishnan,
mukherjee}.

The dynamics of non-interacting quasiperiodic systems has also been
studied recently \cite{neqref1,neqref2,sarkar}. It has been shown
that a driven non-interacting fermionic chain with an incommensurate
Aubry-Andr\'{e} potential and a time-periodic hopping amplitude
exhibits a dynamical transition separating single-particle
delocalized Floquet eigenstates from localized and multifractal
states in the Floquet spectrum. These multifractal Floquet
eigenstates typically occur around the transition frequency.
Moreover, the driven quasiperiodic chain with AA potential, in
contrast to its non-driven counterpart, displays a sharp mobility
edge separating the delocalized and localized or multifractal states
near the transition \cite{sarkar}. However, the fate of these
features remain unclear for the driven AA chain in the presence of
interaction.

In the absence of a drive, an interacting fermion chain with
quasiperiodic potential or random disorder undergoes a transition
between ergodic to MBL phases. The MBL phase, which breaks
ergodicity of an interacting system and hence violates ETH,  has
been extensively studied in the recent past; it is well known that
it leads to qualitatively different long-time behavior of
correlation functions which stems from the absence of ergodicity
\cite{pal,huse,gopalakrishnan,mukherjee}. Moreover, the transition
between the ergodic and MBL phases in 1D interacting systems has
also attracted recent attention. Several studies have shown the
existence of a multifractal phase
\cite{critphase,alet,luitz,garg,abanin2} near the critical point of
MBL transition. The fate of such multifractality in the
thermodynamic limit remains an open question \cite{santos1};
moreover the existence of such states for driven interacting
quasiperiodic systems have not been studied so far.

The study of dynamics in systems near MBL transition have also been
discussed extensively in literature \cite{sid,nag,mukherjee}.
Several such experimental and theoretical studies have been carried
out for periodically driven MBL system both experimentally and
theoretically\cite{abanin,bloch}. Moreover, slow dynamics in the
ergodic phase of a driven MBL in a kicked spin $1/2$ Ising chain
have been reported \cite{Bera}. Recently a many-body critical phase
in the one-dimensional interacting AA model was also predicted; such
a phase turns out to have different properties from both ergodic and
MBL phases \cite{critphase,dassarma}. This seems to suggest that
such quantum system may host three different phases in the
thermodynamic limit \cite{critphase,dassarma}. Unusual correlators
have also been reported in nonequilibrium steady states in strongly
interacting AA model implying several dynamical phases between the
much studied thermal and many body localized phases \cite{swingle}.
However, none of these works have studied the nature of the Floquet
eigenstates near the ergodic to MBL transition point.

In this work, we study a weakly interacting AA model whose hopping
strength is driven by a square pulse protocol. We show that the
Floquet Hamiltonian($H_F$) for such a driven system has extended
ergodic eigenstates at low frequencies; in contrast, they are
many-body localized at large drive frequency. Moreover, $H_F$
supports multifractal eigenstates over a range of driving frequency
$\omega_D$ in the intermediate drive frequency regime. Our results,
within the range of system sizes which could be numerically
accessed, seem to indicate a transition from the ergodic to this
multifractal regime at a critical drive frequency
$\omega_D=\omega_c$ followed by a gradual crossover to the MBL phase
as $\omega_D$ is increased. The multifractal eigenstates that we
obtain possess qualitatively different characteristics from their
ergodic and many-body localized counterparts as is evident from
computation of their IPR and Shannon entropies. We note that such
multifractal eigenstates have been found for disordered many-body
spin and interacting AA Hamiltonians
\cite{luitz,alet,swingle,critphase}; however, to the best of our
knowledge, they have not been reported earlier for a periodically
driven interacting model. Our study therefore provides the
possibility of tuning multifractality of quantum many-body states
using drive frequency.

The other results obtained from our study are as follows. First, we
study the dynamics of representative initial states in different
frequency regimes under the influence of the driven Hamiltonian. We
show that in the intermediate drive frequency regime (which supports
multifractal Floquet eigenstates) they display non-ergodic and
non-MBL behavior. This is evident from the study of both fermion
auto-correlation function and NPR. We find super-exponential decay
of the fermion auto-correlation functions, albeit to a non-zero
value, in this regime; the NPR also shows such intermediate
behavior. Second, we study the half-chain entanglement entropy $S$
which shows a $S \sim a \ln t +b$ growth with $a$ monotonically
decreasing with $\omega_D$ in the intermediate frequency regime.
This growth happens at sufficiently long times; in this time range,
the fermion auto-correlation displays steady oscillation around a
constant value. Third, we find that the half-chain entanglement of
the multifractal eigenstate states show logarithmic growth ($ S \sim
\ln t$) similar to their MBL counterparts; however, the coefficient
of $\ln t$ is controlled by average multifractal dimension of the
eigenstates and can be tuned by $\omega_D$. Fourth, we discuss the
steady state behavior of such a driven system. In particular, we
study the fermion auto-correlation function and fermion density in
the steady state starting from a domain wall initial state (for
which all particles are initially localized to the left half of the
fermion chain). Our results indicate that both the auto-correlation
and the steady state fermion density displays a signature of the
multifractal dimensions and thus can be used to detect multifractal
eigenstates. Fifth, we compute the steady state number entropy
starting from a fermionic product state and discuss its behavior as
a function of the drive frequency. Sixth, we obtain a semi-analytic
Floquet Hamiltonian using a Floquet perturbation theory (FPT) which
reproduces the qualitative features of the driven system obtained
using exact numerics. Our results thus constitutes an analytic
Floquet Hamiltonian which supports multifractal many-body
eigenstates. Finally, we discuss experiments that can test our
theory.

The plan of this paper is as follows. In Sec.\ \ref{hamil} we
discuss the drive protocol that we used throughout our work. Next,
in Sec.\ \ref{phase} we chart out the phase diagram demonstrating
the existence of multifractal Floquet eigenstates for a range of
$\omega_D$. In Sec.\ \ref{dynamics}, we study the short and
intermediate time dynamics of the model. We also discuss the
transport properties of the fermions in the driven interacting AA
chain beginning from a domain wall initial state as well as the
steady state entanglement properties. This is followed by Sec.\
\ref{fptsec} where we use FPT to compute a semi-analytic,
perturbative Floquet Hamiltonian. Finally in Sec.\ \ref{discussion}
we discuss the main results and point out possible experiments which
can test our theory.

\section{The Hamiltonian}
\label{hamil}

 We consider a lattice model that describes 1D
fermions with Aubry-Andr\'{e} (AA) potential and nearest-neighbor
density-density interaction. The Hamiltonian for such a model is
\begin{eqnarray}
H &=& H_{\rm NI} + \sum_j V_{\rm int} \hat n_j \hat n_{j+1}.
\label{ham1}
\end{eqnarray}
where $V_{\rm int}$ is the interaction strength. We consider the
half-filling case for which the ratio of the numbers of fermions $N$
and the lattice sites $L$ is fixed to $N/L = 1/2$. The Hilbert space
dimension is denoted by $\mathcal{N}$. The system is driven by a
periodic square pulse drive protocol described by
\begin{eqnarray}
\mathcal{J}(t) &=& -\mathcal{J}_0,  \quad  t \le T/2  \nonumber\\
&=&\mathcal{J}_0, \quad  t > T/2  \label{sq}
\end{eqnarray}
where $T=2 \pi/\omega_D$ is the time period. In this
study we shall restrict ourselves to the parameter regime $V_{\rm
int} \ll {\mathcal J}_0$. This is done to ensure that the system
remains in the ergodic phase in the quasi-static limit.

In order to study the localization properties of the
driven chain in the Hilbert space, we first need to evaluate the
time evolution operator $U(T,0) = {\mathcal T}_t \exp[-i \int_0^T dt
H(t)/\hbar]$. To this end, we define $H_{\pm} = H[{\mathcal J}=\pm
{\mathcal J}_0]$; the eigenvalues and eigenvectors of $H_{\pm}$ is
given by
\begin{eqnarray}
H_{\pm} |\xi_m^{\pm}\rangle &=& \epsilon_m^{\pm}
|\xi_{m}^{\pm}\rangle.  \label{eign1}
\end{eqnarray}
In terms of these quantities and for the square pulse drive protocol
(Eq.\ \ref{sq}) $U(T,0)$ is given by
\begin{eqnarray}
U(T,0) &=&  e^{- i H_+ T/(2 \hbar)} e^{- i H_- T/(2 \hbar)}
\label{uni} \\
&=& \sum_{p,q} e^{i (\epsilon_p^+ - \epsilon_q^-)T/(2 \hbar)} c_{p
q}^{+-} |\xi_{p}^+\rangle \langle \xi_{q}^-| \nonumber
\end{eqnarray}
where the coefficients $c_{pq}^{+-}= \langle
\xi_{p}^+|\xi_{q}^-\rangle$ denote overlap between the two many body
eigenbasis. In what follows we shall compute $\epsilon_{m}^{\pm}$
and $|\xi_{m}^{\pm}\rangle$ by exact diagonalization (ED). We also
use ED to obtain eigenvalues $\lambda_m$ and eigenvectors
$|\psi_m\rangle$ of $U(T,0)$ . The eigenspectrum of the Floquet
Hamiltonian $H_F$ is found from the relation $U(T,0)= \exp[- i H_F
T/\hbar]$ . Then one can write,
\begin{eqnarray}
U(T,0) &=& \sum_m \lambda_m |\psi_m\rangle \langle \psi_m|, \quad
\lambda_{m} = e^{-i \epsilon^F_{m} T/\hbar} \label{feigen}
\end{eqnarray}
where $\epsilon_m^F$ are the quasienergies which satisfy $H_F
|\psi_m\rangle = \epsilon_m^F |\psi_m\rangle$.

A knowledge of $U(T,0)$ allows us to compute stroboscopic dynamics
starting for an arbitrary initial state $|\psi_{\rm init}\rangle$.
The state at time $t_n=nT$, where $n$ is an integer is given by
\begin{eqnarray}
|\psi(nT)\rangle &=& U(nT,0) |\psi_{\rm init}\rangle = \sum_m
\lambda_m^n  c_m^{\rm init} |\psi_m\rangle \label{wavevol}
\end{eqnarray}
where $c_m^{\rm init} = \langle \psi_m|\psi_{\rm init}\rangle$. Thus
the expectation value of any operator $O$ at stroboscopic times are
given by
\begin{eqnarray}
\langle \psi(nT) |O |\psi(nT)\rangle &=& \sum_{p q} c_p^{ \ast \rm
init} c_q^{\rm init} e^{-i n(\epsilon_q^F -\epsilon_p^F)T/\hbar}
\nonumber\\
&& \times \langle \psi_p | O |\psi_q\rangle \label{opexpec}
\end{eqnarray}
In the steady state, only the terms corresponding to $p=q$ in the
sum (Eq.\ \ref{opexpec}) contribute leading to
\begin{eqnarray}
\langle O \rangle_{\rm steady} &=& \sum_p |c_p^{\rm init}|^2 \langle
\psi_p|O|\psi_p\rangle \label{opsteady}
\end{eqnarray}
We shall use these expressions for study of Floquet dynamics in
subsequent sections.

\section{Phase diagram and the properties of Floquet eigenstates}
\label{phase}

In this section, we shall use the properties of the many-body
Floquet eigenvalues and eigenvectors to study the phase diagram of
the driven chain of length $L$ \cite{das,sarkar} in the presence of
small interaction. First we shall present an exact numerical study
for $L \le 18$ where we have used ED to obtain the exact Floquet
eigenvalues and eigenvectors.

\subsubsection{Inverse participation ratio and fractal dimension}

In order to study the drive induced transition from the ergodic to
the MBL phase in the many-body Fock space basis, we calculate the
inverse participation ratio (IPR) defined as:
\begin{eqnarray} I_m &=&
\sum_{n=1}^\mathcal{N}  |c_{mn}|^4,  \label{ipr1}
\end{eqnarray}
where $c_{mn} = \langle n |\psi_m\rangle $, $|\psi_m\rangle$ is a
Floquet eigenstate, and $|n \rangle$ denotes Fock states in the
number basis. The IPR $I_m \sim \mathcal{N}^{-1(0)}$ in $d=1$ for a
ergodic (MBL) phase and thus acts as a measure of localization of a
many body eigenstate in the Fock space. This property follows from
the fact that a generic many-body ergodic eigenstate of $H_F$ is
expected to have finite overlap with a large number of Fock states;
in contrast, in the MBL phase, it is almost diagonal in the Fock
basis. Thus the behavior of $I_m$ in the Fock space mimics that
inverse participation ratio of single particle Floquet
eigenfunctions function in real space for the non-interacting driven
AA Hamiltonian studied in Ref.\ \onlinecite{sarkar}.

\begin{figure}
\rotatebox{0}{\includegraphics*[width= 0.48 \linewidth]{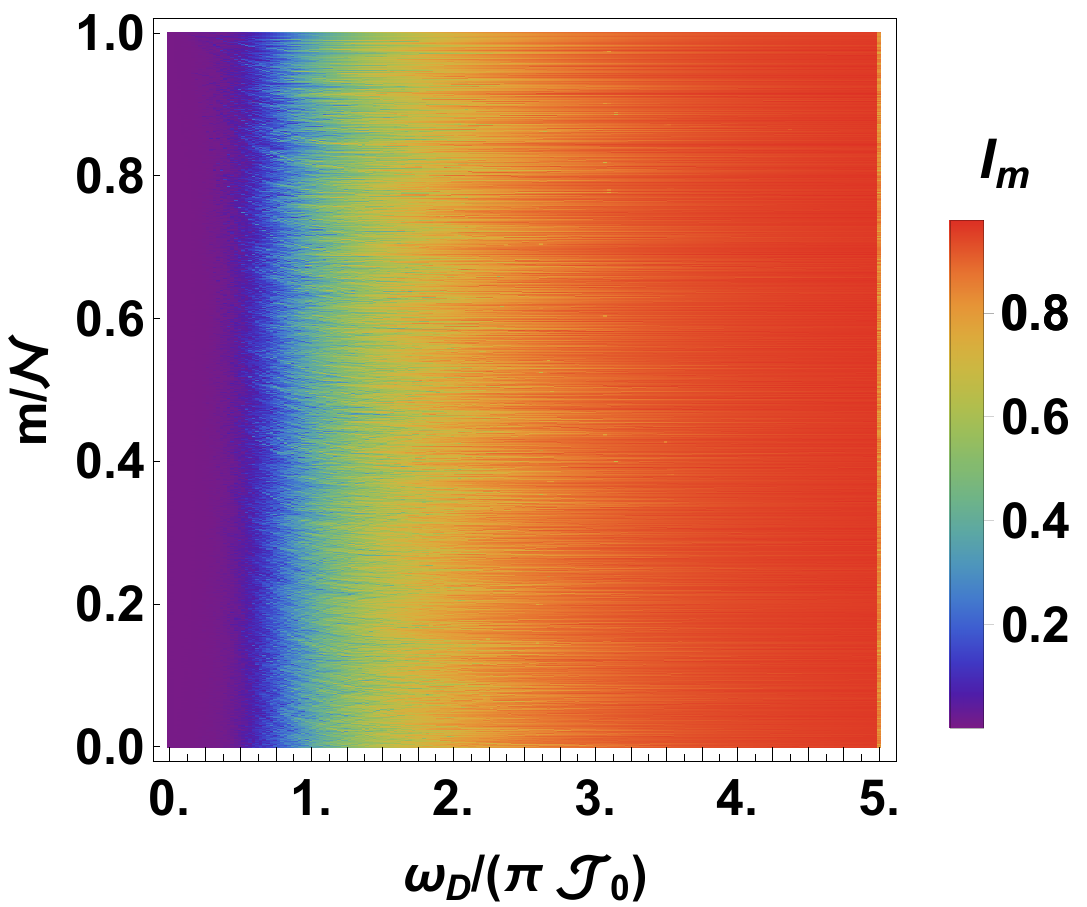}}
\rotatebox{0}{\includegraphics*[width= 0.50 \linewidth]{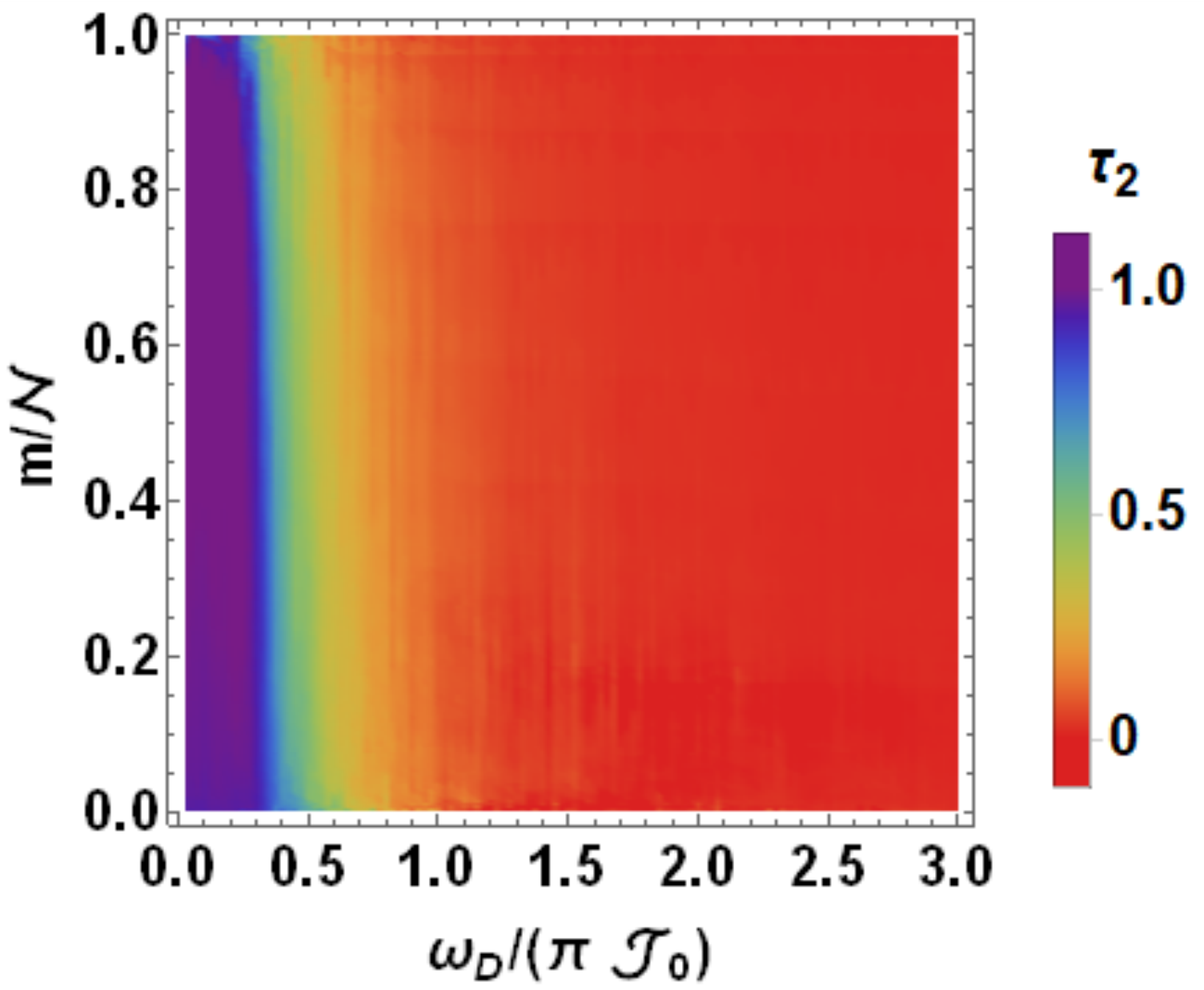}}
\rotatebox{0}{\includegraphics*[width= 0.49 \linewidth]{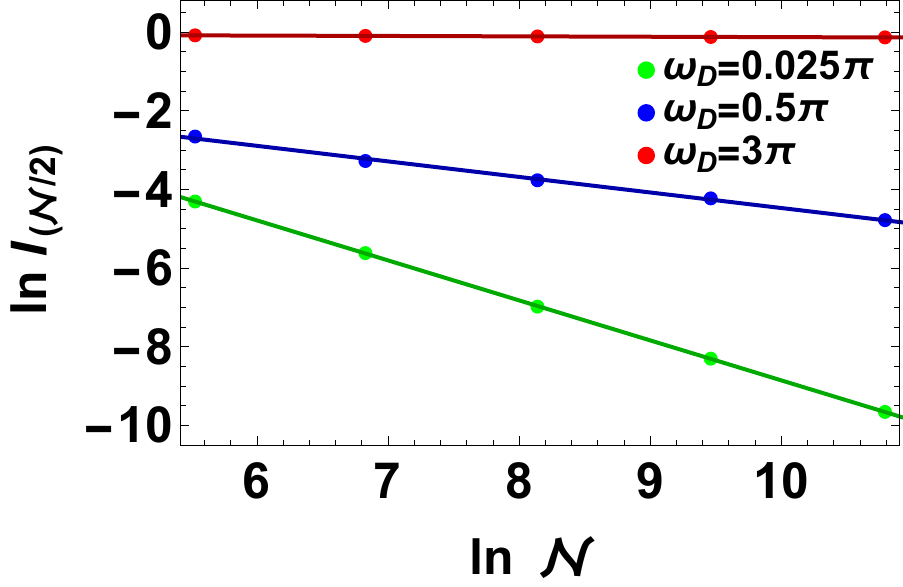}}
\rotatebox{0}{\includegraphics*[width= 0.49 \linewidth]{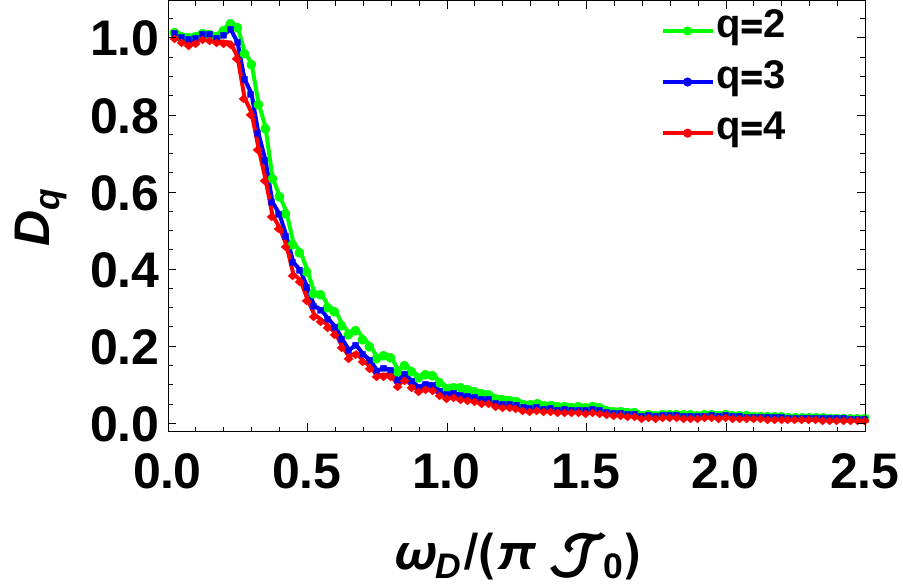}}
\caption{Top Left Panel: Plot of $I_m$ as a function of the
normalized many-body eigenfunction index $m/\mathcal{N}$ and
$\omega_D/(\pi \mathcal{J}_0)$ showing the localized/delocalized
nature of the Floquet eigenstates $|\psi_m\rangle$ for $L=14$. Top
Right panel: Plot of $\tau_2$ as a function of $m/\mathcal{N}$
(after sorting in increasing order of $I_m$) and $\omega_D/(\pi
\mathcal{J}_0)$ showing the presence of delocalized states for $
\omega_D/(\pi \mathcal{J}_0)\le 0.4$, multifractal states for $0.4
\le \omega_D/(\pi \mathcal{J}_0)\le 1.5$ and fully localized states
for $ \omega_D/(\pi \mathcal{J}_0) > 1.5$. The system sizes used for
extracting $\tau_2$ are $L=10, \cdots, 18$ in steps of $2$. Bottom
Left panel: Plot for $\ln I_m$ vs $\ln L$ used for extracting
$\tau_2$ for several representative frequencies for the state
corresponding to $m/{\mathcal N}=0.5$. The behavior of perfectly
delocalized (green dots at $\omega_D/(\pi \mathcal{J}_0)=0.025 $)
and localized (red dots, $\omega_D/(\pi \mathcal{J}_0) =3$) can be
distinguished from that of a multifractal states (blue dots
$\omega_D/(\pi \mathcal{J}_0)=0.5$). Bottom Right Panel: Plot of
$D_q$ as a function of $\omega_D/(\pi \mathcal{J}_0)$ for
$m/\mathcal{N}=0.5$. We have set $\mathcal{J}_0=1$,
$V_0/\mathcal{J}_0=0.05$ , $V_{\rm int}/\mathcal{J}_0=0.025$ ,
scaled all energies and frequencies in units of $\mathcal{J}_0$
(with $\hbar$ set to unity). See text for details.} \label{fig1}
\end{figure}

The analysis of $I_m$ leads to the phase diagram shown in top left
panel of Fig.\ \ref{fig1}, where $I_m$ is plotted as a function of
eigenvector index $m/{\mathcal N}$ and $\omega_D$. The plot shows
that the driven AA model with interaction exhibits a transition from
the ergodic to the MBL phase. For low drive frequencies $\hbar
\omega_D/(\pi \mathcal{J}_0) < 0.4$, all Floquet eigenstates are
ergodic with $I_m \sim (1/\mathcal{N})$. A transition from ergodic
eigenstates to a phase where the eigenstates states with $ 0 < I_m <
1 $ occur around $\hbar \omega_D/(\mathcal{J}_0 \pi) \sim 0.4$.
These eigenstates (which have $0 < I_m < 1$) persists for a wide
range of frequencies $0.4 \le \hbar \omega_D/(\pi {\mathcal J}_0)
\le 1.5$. For $\hbar \omega_D/(\pi {\mathcal J}_0) \gg 1.5$, the
Floquet eigenstates become completely localized ($I_m \simeq 1$)
signifying the onset of the MBL phase.

To study the nature of states having $0 < I_m < 1 $, we compute
\begin{eqnarray}
I_m^{(q)} &=&  \sum_{n=1}^{{\mathcal N}}
|c_{mn}|^{2q} \label{iqdef}
\end{eqnarray}
where $I_m \equiv I_m^{(2)}$. It is well known $I_m^{(q)} \sim
{\mathcal N}^{-\tau_q}$, where the exponent $\tau_q$ is related to
the fractal dimension $D_q$ by $D_q= \tau_q/(q-1)$. We note that for
MBL states, we expect $D_q=0$ whereas for ergodic states $D_q=1$.
The intermediate $q$ dependent values of $D_q$, that is $D_q=
\tau_q/(q-1)$, signifies multifractality while $D_q$ is independent
of $q$ for a fractal eigenstate.

To analyze the nature of the Floquet eigenstates further, we first
plot $\tau_2$ as a function of eigenvector index $m/{\mathcal N}$
and $\omega_D$ in the top right panel of Fig.\ \ref{fig1}. From this
plot, we find the presence of ergodic and MBL states for low ($\hbar
\omega_D/(\pi \mathcal J_0) <0.4$) and high ($\hbar \omega_D/(\pi
\mathcal J_0) > 1.5 $) drive frequencies respectively. In between,
one finds state with $0 \le \tau_2 \le 1$ signifying their
non-ergodic and non-MBL nature. We note that for this plot we sort
the eigenstates in increasing value of $I_m$. Thus we find that the
states which had $ 0 \le I_m \le 1 $ also have $0 \le \tau_2 \le 1$;
these states are natural candidate for multifractal Floquet
eigenstates.

In what follows, we extract $\tau_2$ from the plot of $ \ln I_m $
versus $ \ln \mathcal{N} $ as shown in the bottom left panel of
Fig.\ \ref{fig1} for $m/{\mathcal N}=0.5$. For $\hbar \omega_D/(\pi
\mathcal{J}_0)=3$ the state is many-body localized and we have
$\tau_2 \sim 0 $ as evident from the flat red line in the bottom left panel
of Fig.\ \ref{fig1}. In contrast, at $\hbar \omega_D/(\pi
\mathcal{J}_0)=0.025 $ we have $\tau_2=1$ (green line in bottom left panel of
Fig.\ \ref{fig1}) signifying the ergodicity of the state. In
between, at $\hbar \omega_D/(\pi \mathcal{J}_0) =0.5$ , $\tau_2=0.4$
(blue line in bottom left panel of Fig.\ \ref{fig1}) indicating the
presence of non-ergodic and non-MBL nature of the state.

The plot of the multifractal dimension $D_q$ is shown in bottom
right panel of Fig.\ \ref{fig1} for states corresponding to
$m/{\mathcal N}=1/2$. For all points in these plots, $D_q$ is
obtained from values of $\tau_q$ that are, in turn, extracted from
the corresponding plots of $\ln I_m$ vs $\ln {\mathcal N}$. From the
plot, we find that for $ 0.35 \le \hbar \omega_D/(\pi \mathcal{J}_0)
\le 1.5$, $0\le D_q \le 1$; this indicates the presence of
multifractal states in the spectrum. Other states with different
$m/{\mathcal N}$ also show similar features. The behavior of $D_q$
shown in the bottom right panel of Fig.\ \ref{fig1} indicates that
the driven fermion chain exits the ergodic phase for $\hbar
\omega_D/(\pi \mathcal{J}_0) \simeq 0.4$. We note here that our
numerical analysis shows that $D_q$ is almost independent of $q$ for
$q\le 4$; this indicates the possibility of fractal nature of these
states. However, ascertaining this property would require
computation of $\tau_q$ for all $q$ and we do not attempt this in
the present work.

\subsubsection{Shannon Entropy}

To further establish the presence of the ergodic , multifractal and
MBL phases as a function of frequency and to find out the nature of
the transition between them, we study the Shannon entropy of the
Floquet eigenstates.  The Shannon entropy of the $m^{\rm th}$
Floquet eigenstate is given by
\begin{eqnarray} S_m &=& -\sum_m |c_{mn}|^2 \ln |c_{mn}|^2, \quad S
= \frac{1}{\mathcal{N}} \sum_m S_m  \nonumber\\ \label{shannon1}
\end{eqnarray}
where $S$ is the mean entropy. We note that for $\hbar \omega_D/(\pi
\mathcal{J}_0) \gg 1$, $c_{mn} \simeq \delta_{mn}$ leading to $S_m
\simeq 0$; thus $S \to 0$ indicates many-body localized
eigenfunctions. In contrast for $\hbar \omega_D/ (\pi \mathcal{J}_0)
\ll 1$ when all Floquet eigenstates are ergodic, $c_{mn} \simeq
1/\sqrt{\mathcal{N}}$ for all $m$ leading to maximum entropy of $S =
S_{\rm max} \simeq \ln \mathcal{N}$.

\begin{figure}
\rotatebox{0}{\includegraphics*[width= 0.48
\linewidth]{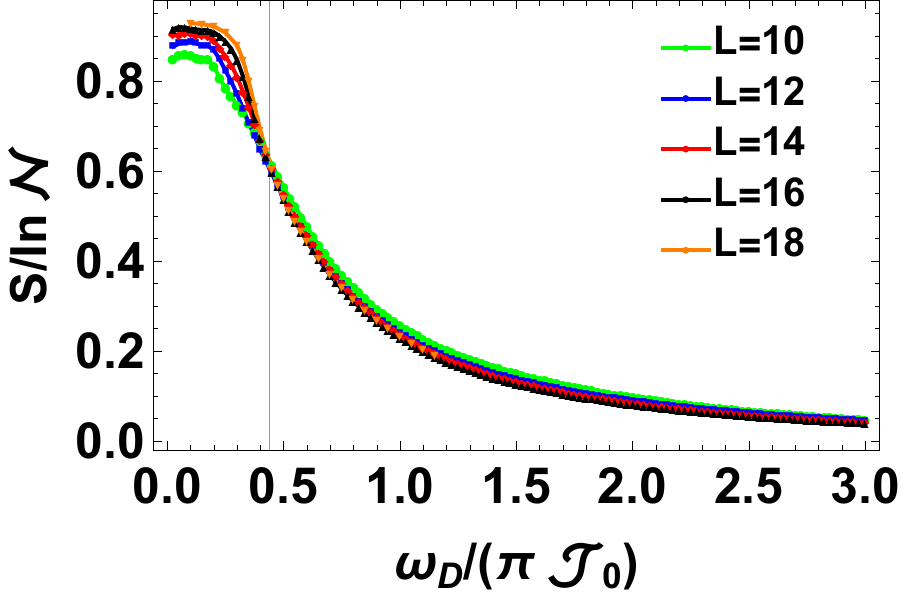}} \rotatebox{0}{\includegraphics*[width=
0.48 \linewidth]{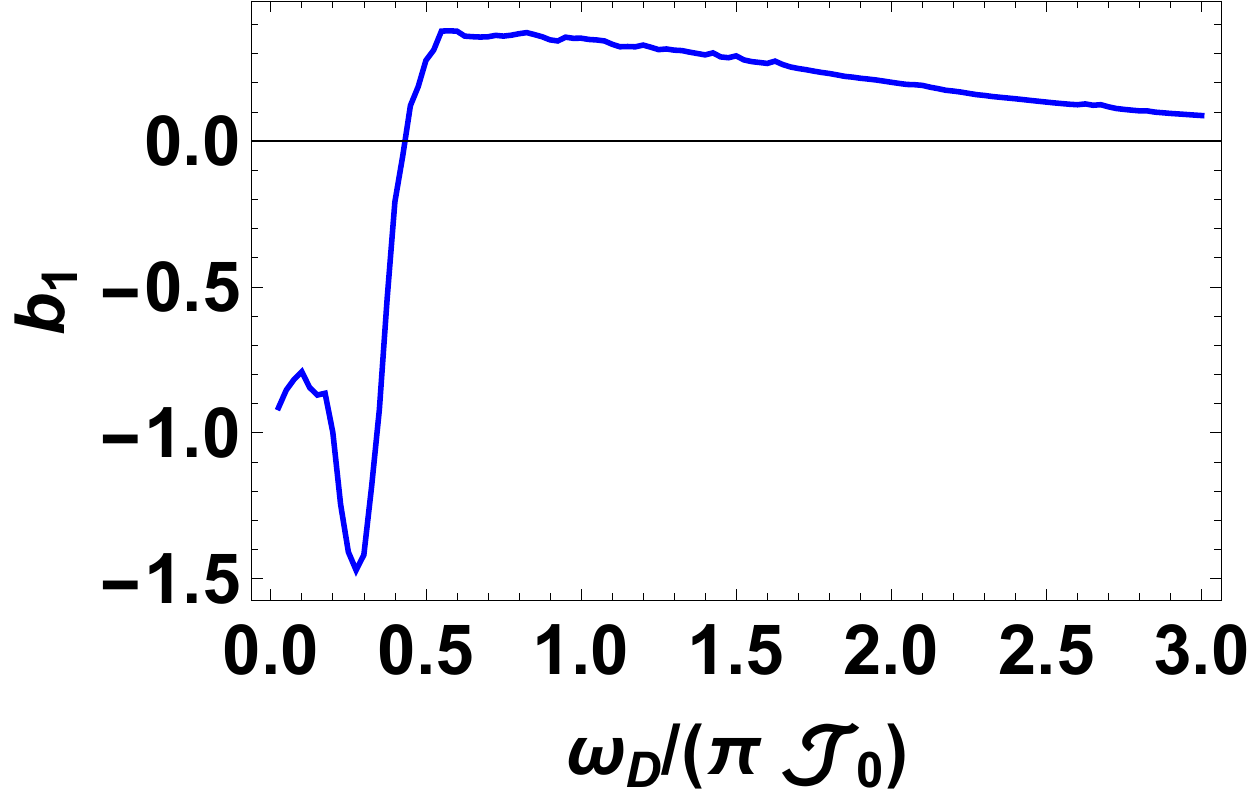}} \caption{ Left panel: Plot of $S /\ln
\mathcal{N}$ as a function of $\omega_D/(\pi\mathcal{J}_0)$ showing
the presence of an ergodic to MBL transition at $\omega_D/(\pi
\mathcal{J}_0) \sim 0.44$. Right panel: Plot of $b_1$ as a function
of $\omega_D/(\pi\mathcal{J}_0)$ showing $b_1 \le 0$ in the ergodic
phase whereas $b_1 > 0$ in the MBL phase . The plot at right panel
shows that $b$ changes sign at $\omega_D/(\pi \mathcal{J}_0) \sim
0.45$ suggesting a critical point. All other parameters are same as
in Fig.\ \ref{fig1}. See text for details.} \label{fig2}
\end{figure}

The left panel of Fig.\ \ref{fig2} shows the plot of Shannon entropy
normalized by $\ln \mathcal{N}$ as a function of $\hbar
\omega_D/\mathcal{J}_0$ for different system sizes. Note that the
plot for different system sizes cross each other around $\hbar
\omega_D/\mathcal{J}_0 \sim 0.43 \pi $; this seems to indicate a
transition between the ergodic and multifractal phases. To
understand this feature further, we note that the functional form of
$S$ can be written as \cite{alet}
\begin{eqnarray}
S &=& D_1 \ln \mathcal{N} + b_1 \label{enteq}
\end{eqnarray}
where $D_1$ is the fractal dimension. It is known that $b_1$ is
expected to change sign at the transition from ergodic to MBL phase.
This results in a crossing point between the curves of $S$ of
different sizes at the critical frequency \cite{alet}. The plot of
$b_1$ is shown in the right panel of Fig.\ \ref{fig2}. It indicates
that delocalized phase $b_1 < 0$ whereas in the MBL phase $b_1 > 0$.

It is also instructive to study the fluctuations of entanglement
entropy \cite{abanin}, as they have been shown to provide a useful
probe of the delocalization to MBL transition. The fluctuations of S
is defined as,
\begin{eqnarray}
\Delta S = \sqrt{ \langle (S - \langle S \rangle )^2 \rangle }
\end{eqnarray}
It is known that $\Delta S$ is small deep inside both the ergodic
and MBL phases. In the ergodic phase, all Floquet eigenstates are
highly entangled with $S=S_{\rm max}$. Thus the system exhibits
small fluctuation around this value. In the MBL phase $S$ follows an
area law and is hence small (compared to that in the ergodic phase
where it follows volume law). In addition, since all states have low
$S$, the fluctuations are small. In contrast, at the transition $S$
has a broad distribution leading to maximal value of $\Delta S$.
Thus, the delocalization to MBL transition can be detected by the
location of the peak in $\Delta S$.

\begin{figure}
\rotatebox{0}{\includegraphics*[width= 0.8 \linewidth]{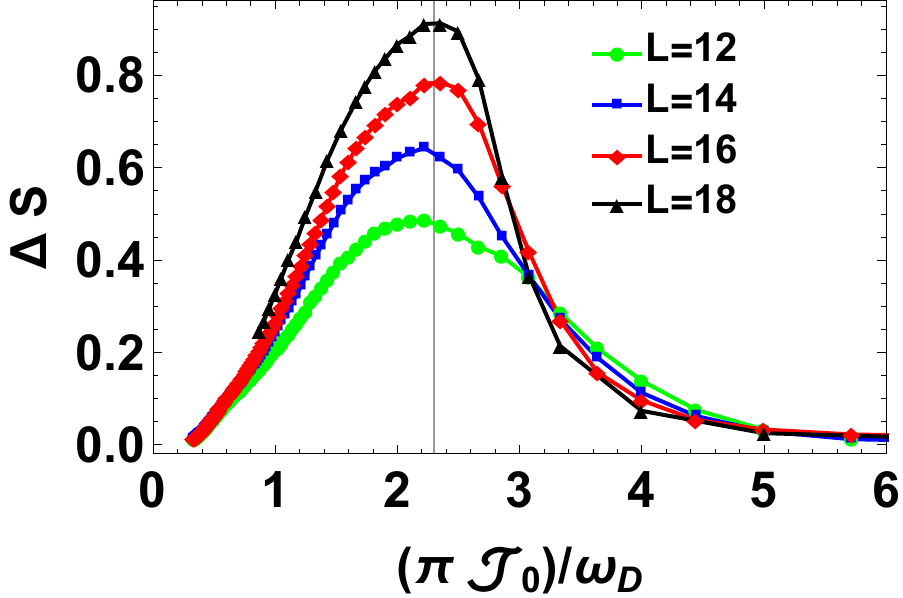}}
\caption{ Plot of $\Delta S$ as a function of
$(\pi\mathcal{J}_0)/\omega_D$ showing that the transition from
ergodic to the multifractal phase takes place at  $\pi
\mathcal{J}_0/\omega_D \sim 2.3 $ or $\omega_D/\mathcal{J}_0 \sim
0.434 \pi$. All other parameters are same as in Fig.\ \ref{fig1}.
See text for details.} \label{fig3}
\end{figure}

Fig.\ \ref{fig3} shows the plot of $\Delta S$ as a function of $\pi
\mathcal{J}_0 / \hbar \omega_D$ and confirms that such a peak
appears at $(\pi \mathcal{J}_0)/\hbar \omega_D \sim 2.3$. We note
that the peak gets sharper with increasing system size with a slight
shift towards higher $T$; this indicates that drive may possibly
induce a transition between ergodic and non-ergodic(multifractal)
states which shall survive for larger $L$. Combining the results of
$D_q$ and $\Delta S$, we seem to find that within the finite system
sizes that we can access, there is possible transition around $\hbar
\omega_D/(\pi \mathcal{J}_0) \simeq 1/2.3= 0.434$. A more definite
characterization of this possible transition would require access to
larger system size which is outside the scope of the present work.

\section{Quantum Dynamics}
\label{dynamics}

In this section we discuss dynamical signatures for the multifractal
states in the region of intermediate frequencies. We divide this
section to study three different timescales, namely, short,
intermediate and long time steady state. We analyze the behavior of
different correlation functions and entropies in different regimes.
This analysis is expected to be useful from an experimental
standpoint since achieving short-time coherent dynamics is easier in
experiments. Thus the signatures of multifractal states visible in
those time-scales, if any, is much easier to detect experimentally.

\subsection{Short-time Dynamics}
\label{shortdynamics}

In this section we shall study the evolution of a  product initial
state in the basis of $H$ in the short time regime, $n_0 <10^2$
cycles. We look for possible signatures of multifractal eigenstates
of $H_F$ in dynamics which are different from the dynamics induced
by either ergodic and MBL eigenstates. Unless otherwise mentioned,
all the quantities in this section are calculated by averaging over
$N_0$ product initial states. We choose $N_0= 500$ for sizes
$L=12,14$, $N_0=100$ for sizes $L=16,18$, and $N_0=18[20]$ for
$L=20[22]$. We have chosen $N_0$ such that the error bars are
smaller than the size of points. To evolve the system, we have used
standard Krylov subspace techniques. \cite{krylov}.

We show several features which appear to be intermediate between
ergodicity and MBL in the range of frequencies where multifractality
appears in the Floquet spectrum. These features are present for all
$L$ studied here and their presence is therefore expected to be
independent of $L$ at least for the range of system sizes studied in
this work. It is important to note that while the eigenstate
properties are calculated for size $L=10-18$, the absence of
significant deviation in results for all $L \le 22$ seems to justify
our claim of the presence of a non-ergodic multifractal regime for
larger system sizes than what can be accessed by ED.

\subsubsection{Auto-correlation function}

In order to distinguish between ergodic, multifractal and MBL
phases, we resort to the measurement of temporal auto-correlation
function. The auto-correlation function is a measure of retention of
memory of system's initial state \cite{rafael} and is given by
\begin{eqnarray}
\mathcal{A}_j (t) = \left( 2 \langle \hat n_j  (t) \rangle -1
\right)\left( 2 \langle \hat n_j  (0) \rangle -1 \right)
\end{eqnarray}
where $\mathcal{A}_j (t)$ is the temporal auto-correlator at site
$j$ and time $t=n_0T$ and $\langle n_j(t) \rangle = \langle
U^{\dagger}(n_0 T,0) n_j(0) U(n_0 T,0)\rangle$ is the expectation
value of fermion number operator at site $j$ and time $t= n_0 T$. We
average this single site operator over different sites and over different random product 
initial states
\begin{eqnarray}
\mathcal{A}(t) = \frac{1}{L} \left[\sum_{j=1}^L \mathcal{A}_j
(t)\right]
\end{eqnarray}
where the square brackets indicate initial state averaging.
\begin{figure}[H]
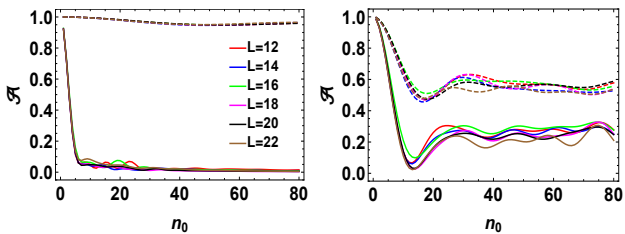

\centering
{\ing[width=0.47\linewidth,height=0.35\columnwidth]{{short_auto1}.pdf}}
\centering
{\ing[width=0.47\linewidth,height=0.35\columnwidth]{{short_auto2}.pdf}}
\caption{Plot of the temporal auto-correlation function as a
function of number of cycles $n_0$. Left panel: Solid lines
correspond to $\omega_D/(\pi \mathcal{J}_0)= 0.1$ (ergodic phase)
while the dashed lines pertains to $\omega_D/(\pi \mathcal{J}_0)=
2.5$ (MBL phase). Right panel: Solid lines correspond to
$\omega_D/(\pi \mathcal{J}_0)= 0.4$ near the transition from the
ergodic to the multifractal regime while the dashed lines pertains
to $\omega_D/(\pi \mathcal{J}_0)= 0.7$ (multifractal regime). All
other parameters are same as in Fig.\ \ref{fig1}. See text for
details.} \label{auto}
\end{figure}

We can distinguish between the three phases using the $\mathcal{A}$
versus $n_0$ plot as indicated in Fig.\ \ref{auto} where $n_0$
denotes the number of drive cycles. It is known that for short range
systems that ${\mathcal A}$ displays exponential decay in the
ergodic phase. This behavior is seen at low frequencies $\hbar
\omega_D/(\pi \mathcal{J}_0)= 0.1$, as shown in the left panel of
Fig.\ \ref{auto} (solid lines). The temporal auto-correlation
function reduces to $1/L$ rapidly over a short interval of time. Due
to the long range nature of the Floquet Hamiltonian, the decay
deviates a bit from the usual exponential decay. In contrast, in the
MBL phase at $\hbar \omega_D/(\pi {\mathcal J}_0) > 1.5$, the system
is supposed to retain the memory of it's initial state for a very
long time. For short timescales ($n_0 \le 100$) studied in this
section the auto-correlation does not decay significantly. This
feature is seen at a high frequency of $\hbar \omega_D/(\pi
\mathcal{J}_0)= 2.5$ in the left panel of Fig.\ \ref{auto} (dashed
lines).

\begin{figure}[H]
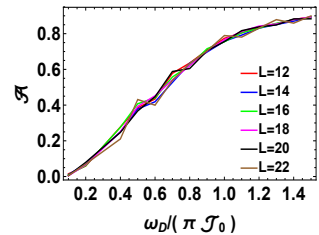

\centering
{\ing[width=0.47\linewidth,height=0.35\columnwidth]{{autovsfreq}.pdf}}
\caption{Plot of the temporal auto-correlation function as a
function of $\omega_D/(\pi \mathcal{J}_0)$ after a fixed number of
cycles $n_0=80$ for different $L$. All other parameters are same as
in Fig.\ \ref{fig1}. See text for details.} \label{Avsfreq}
\end{figure}

For the region of multifractal frequencies, it is not immediately
clear how $\mathcal{A}$ should behave as the wavefunctions are
extended but the system cannot be called ergodic. Our numerical
result in the right panel of Fig.\ \ref{auto} shows that for $\hbar
\omega_D/(\pi \mathcal{J}_0)=0.4$ (solid line) and $0.7$ (dashed
line), $\mathcal{A}$ shows a decay initially but then oscillates
around a value which is intermediate to $1/L$ and $1$. The value of
$\omega_D$ which controls the multifractal dimensions of the
eigenstates of $H_F$ has effect on both the rate of the decay of
${\mathcal A}$ and its final value. This will be analyzed in details
in subsequent sections. We note that for all $L$ considered,there is no
significant finite size effect as can be seen from Fig.
\ref{Avsfreq}. Thus the behavior of ${\mathcal
A}(n_0)$ shown by Figs.\ \ref{auto} and \ref{Avsfreq} definitely
points to the presence of a non-ergodic and non-MBL phase in the
region of intermediate drive frequencies which supports multifractal
eigenfunctions of $H_F$.

\subsubsection{Normalized Participation Ratio}

One of the most common quantity to characterize delocalization-MBL
transition is the measurement of the normalized participation ratio
(NPR) which provides information about the volume of phase space
explored by the system during dynamics \cite{rafael}. The NPR is
defined as
\begin{eqnarray}
N_p^{(m)}(t) &=& \frac{1}{ P_m(t)  \mathcal{N}}, \quad
\zeta^{(m)}(t)= \ln N_p^{(m)}(t) \label{nprdef}
\end{eqnarray}
where $\mathcal{N}$ is the Hilbert space dimension, $t \in n_0 T$,
and  $P_m(t)=\sum_n |d_n(t)|^{2m}$ is the dynamical IPR. In Eq.\
\ref{nprdef}, $d_n(t)=\langle\chi_n|\psi^{\prime}\rangle$ where
$|\psi^{\prime}\rangle=U(n_0 T,0)|\psi_{\rm init}\rangle$ and
$|\chi_n\rangle$ are the computational basis states. Here we choose
several initial states $|\psi_{\rm init}\rangle$ and average
$\zeta^{(m)}(t)$ over all such initial states. Also, for the rest of
this section, we shall denote $\zeta(t) =
[\zeta^{(2)}(t)]$ for clarity.

We note that $\zeta$(t) denotes the fraction of the configuration
space that the system explores. In the delocalized phase we expect
$\zeta(t)$ to be independent of $L$ and to reach the maximum value
of zero when the system is uniformly ergodic. In the high frequency
regime, when the system is in a MBL phase $\zeta(t)$ varies with
$L$.

\begin{figure}[H]
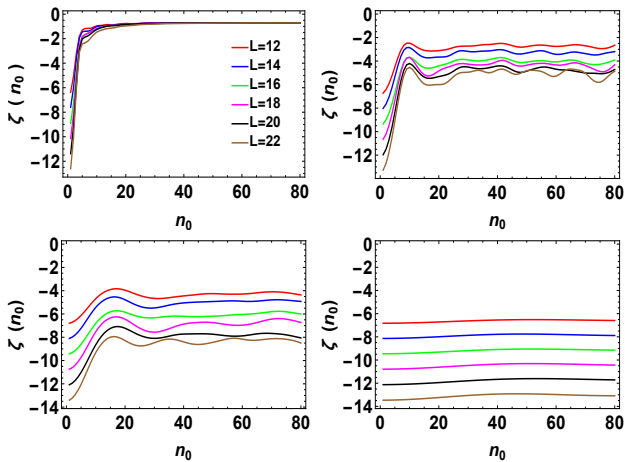

\centering
{\ing[width=0.47\linewidth,height=0.35\columnwidth]{{frachil_omega0.1}.pdf}}
\centering
{\ing[width=0.47\linewidth,height=0.35\columnwidth]{{frachil_omega0.4}.pdf}}
\centering
{\ing[width=0.47\linewidth,height=0.35\columnwidth]{{frachil_omega0.7}.pdf}}
\centering
{\ing[width=0.47\linewidth,height=0.35\columnwidth]{{frachil_omega2.5}.pdf}}
\caption{Plot of $\zeta$ as a function of $n_0$ for $\omega_D/(\pi
\mathcal{J}_0)= 0.1$ corresponding to the ergodic regime (top left
panel) and $\omega_D/(\pi \mathcal{J}_0)= 0.4$ near the transition
from the ergodic to the multifractal regime (top right panel). The
bottom left panel corresponds to $\omega_D/(\pi \mathcal{J}_0)= 0.7$
(multifractal regime) and the bottom right panel pertains to
$\omega_D/(\pi \mathcal{J}_0)= 2.5$ (MBL phase). All other
parameters are same as in Fig.\ \ref{fig1}. See text for details.}
\label{hil}
\end{figure}
Fig.\ \ref{hil} shows the plot of $\zeta$ as a function of number of
drive cycles $n_0$ for different frequencies. For $\hbar
\omega_D/(\pi \mathcal{J}_0)= 0.1$, in the delocalized region, as
shown in the top left panel of Fig.\ \ref{hil}, $P_m \to
1/\mathcal{N}$ and hence $\zeta \to 0$. The plots for
various system sizes therefore converges together to a
$L$-independent near-zero value signifying ergodicity. In contrast,
in the localized region $I_m \to 1$ and hence $\zeta \to - \ln
\mathcal{N}$. Thus $\zeta$ varies with $L$ as shown in the bottom
right panel of Fig.\ \ref{hil} for $\hbar \omega_D/(\pi
\mathcal{J}_0)= 2.5$.

In the intermediate frequency regime, as shown for $ \hbar
\omega_D/(\pi \mathcal{J}_0)= 0.4 (0.7)$ in top right(bottom left)
panel of Fig.\ \ref{hil}, we find $-\ln \mathcal{N} < \zeta < 0 $.
This behavior, seen throughout the intermediate frequency range,
shows that the phase space exploration originating from multifractal
eigenstates of $H_F$ is faster than that due MBL eigenstates but
slower than the ergodic ones. If the frequency is closer to the
ergodic region where $\tau_2$ is closer to unity, $\zeta$ for
different $L$ converge with increasing $\omega_D$. However, this
behavior is different from the complete convergence found for
ergodic Floquet eigenstates. As the drive frequency is increased,
$\tau_2$ obtained from eigenstates of $H_F$ decreases. This leads to
an increased separation (at the time scales studied here) of $\zeta$
for different $L$; moreover, the magnitude of change of $\zeta$
becomes smaller compared to the initial value. This intermediate
behavior of $\zeta$ also points to the presence of non-ergodic and
non-MBL states in agreement that seen from analysis of
$\mathcal{A}$.

\subsubsection{Entanglement Entropy}

In this section, we introduce half-chain von-Neumann entanglement
entropy \cite{roop}, denoted by $S_{vN}$ for the driven chain.
$S_{vN}$ can be defined in terms of the reduced density matrix
$\rho_A$ of the chain after $n_0$ drive cycles. This is computed
 by tracing out the density matrix $\rho= |\psi (n_0
T)\rangle \langle \psi (n_0 T) |$ (where $|\psi(n_0 T)\rangle =
U(n_0 T,0) |\psi_{\rm init}\rangle$) over half the chain. In terms
of this reduced density matrix $\rho_A$, one then obtains
\begin{eqnarray}
S_{vN} (n_0 T) = - {\rm Tr} [ \rho_A(n_0 T) \ln \rho_A(n_0 T) ].
\end{eqnarray}
For MBL states, $S_{vN} (n_0 T) \sim \ln n_0$ if one starts from a
homogeneous initial state\cite{numberentrp}; in contrast $S_{vN}
\sim n_0$ for ergodic states. For systems that support multifractal
states there have some studies of how multifractal dimensions of
states determine the  entanglement entropy\cite{mfracentr}.
For the present case, we note that the driven interacting
fermion chain supports Floquet eigenstates of different multifractal
dimensions controlled by $\omega_D$. Moreover, for a fixed
$\omega_D$, it supports a spectrum of $\tau_2$ (Fig. \ref{fig2}).
This suggests that the behavior of $S_{vN}$ for such eigenstates may
be unconventional. However, one requires to probe into much larger
time scales than that discussed in this subsection to probe the
precise $n_0$ dependence of $S_{vN}$. This will be addressed in the
next subsection where we discuss intermediate time behavior.
Moreover, the measurement of $S_{vN}$ in experiments is a difficult
task. In contrast, very recently, a different kind of entropy called
number entropy has been shown to be experimentally measurable
\cite{numberentr}.  We shall therefore study the short time behavior
of the number entropy in the rest of this subsection.

In systems where the total particle number is conserved (which holds
for the present case), the von-Neumann Entropy can be split into two
parts\cite{numberentr}:
\begin{eqnarray}
S_{vN}=S_c + S_N
\end{eqnarray}
where $S_c$ is the configuration entropy and $S_N$ is the number
entropy. $S_N$ characterizes particle number fluctuation in the
subsystem under consideration and is defined as,
\begin{equation}
S_N = - \sum_n p(n) \, \ln p(n)
\end{equation}
where $p(n)$ is the probability of finding $n$
fermions within the subsystem (half chain for our case).
\begin{figure}[H]
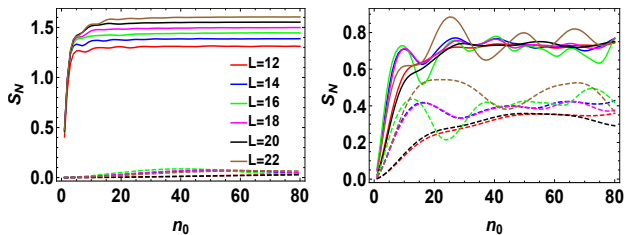

\centering
{\ing[width=0.47\linewidth,height=0.35\columnwidth]{{short_SN1}.pdf}}
\centering
{\ing[width=0.47\linewidth,height=0.35\columnwidth]{{short_SN2}.pdf}}
\caption{Plot of $S_{N}$ as a function of $n_0$. The solid lines in
the left panel correspond to the ergodic phase ($\omega_D/(\pi
\mathcal{J}_0)= 0.1$) while the dashed lines pertains to the MBL
phase ($\omega_D/(\pi \mathcal{J}_0)= 2.5$). The solid lines in the
right panel correspond to $\omega_D/(\pi \mathcal{J}_0)= 0.4$ (near
the transition from the ergodic to the multifractal regime) and the
dashed lines pertains to the multifractal phase ($\omega_D/(\pi
\mathcal{J}_0)= 0.7$). All other parameters are same as in Fig.\
\ref{fig1}. See text for details.} \label{NE}
\end{figure}
It is expected that in ergodic phase, $S_N \sim \ln t$. Moreover, it
has been numerically shown recently that in MBL phase (in contrast
to the previously prediction of system size independent saturation),
$S_N \sim \ln \ln t$. \cite{numberentrp}. The study of the temporal
dependence of $S_N$ to confirm such $\ln t$ (or $\ln \ln t$)
behavior will be taken up in the next subsection. Here, we plot
$S_N$ as a function of number of drive cycles $n_0$ for different
representative drive frequencies at short-times and discuss whether
there are any markers of the multifractal phase. The solid lines in
the left panel of Fig.\ \ref{NE} shows the behavior of $S_N$ for
$\hbar \omega_D/(\pi \mathcal{J}_0)= 0.1$ (ergodic regime); here
$S_N$ seems to display a fast logarithmic growth before it saturates
to an $L$ dependent value. The dashed lines in the left panel of
Fig.\ \ref{NE} shows that for $\hbar \omega_D/(\pi {\mathcal J}_0)=
2.5$ (MBL phase) $S_N$ is almost a constant.

In the multifractal region , for $\hbar \omega_D/(\pi
\mathcal{J}_0)= 0.4$ (solid line) and $0.7$ (dashed line) as shown
in the right panel of Fig.\ \ref{NE}, $S_N$ displays a
sub-logarithmic growth followed by oscillations around a constant
value. The amplitude of these oscillations increases with $L$ within
the range of system sizes studied here. These features distinguish
the multifractal phase from both the ergodic and the MBL  phases and
shows that $S_N$ carries signature of the multifractal phase
realized at intermediate drive frequencies.

\begin{figure}[H]
\centering
{\ing[width=0.47\linewidth,height=0.35\columnwidth]{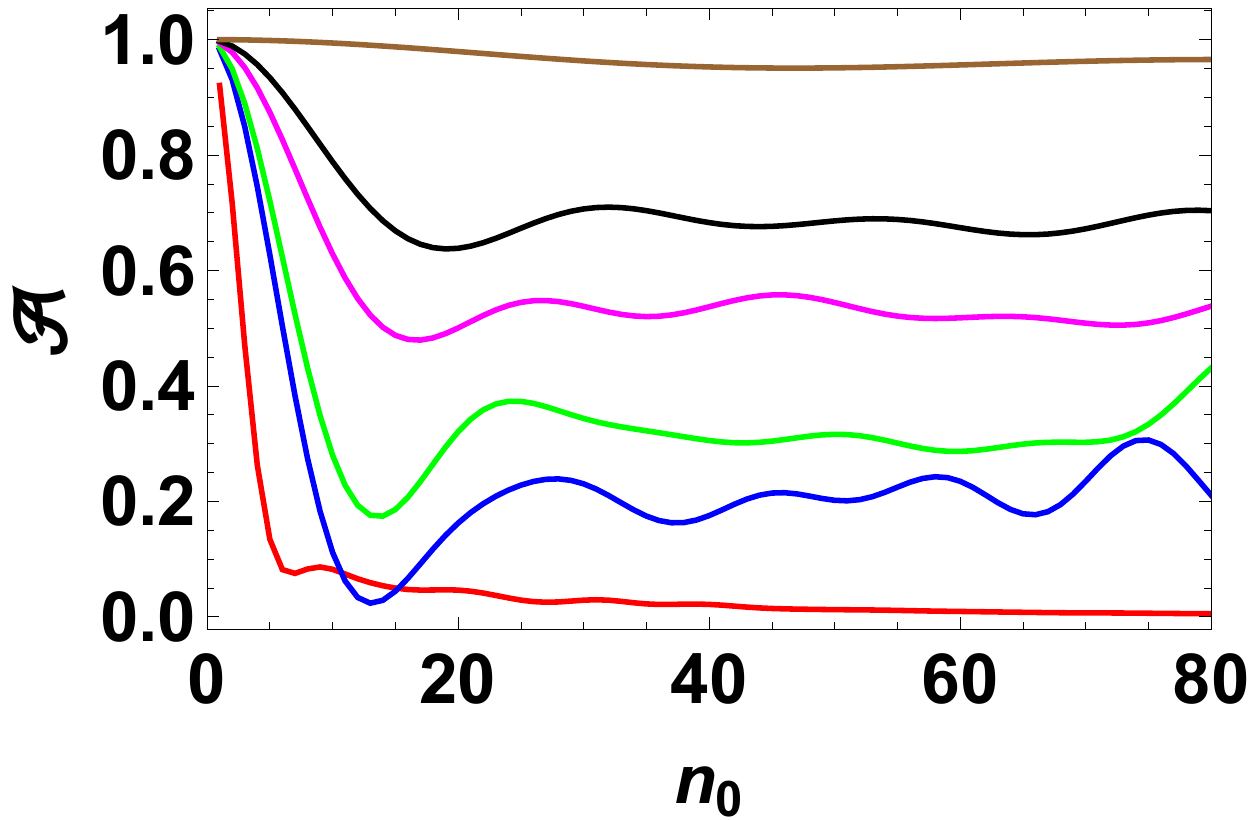}}
\centering
{\ing[width=0.47\linewidth,height=0.35\columnwidth]{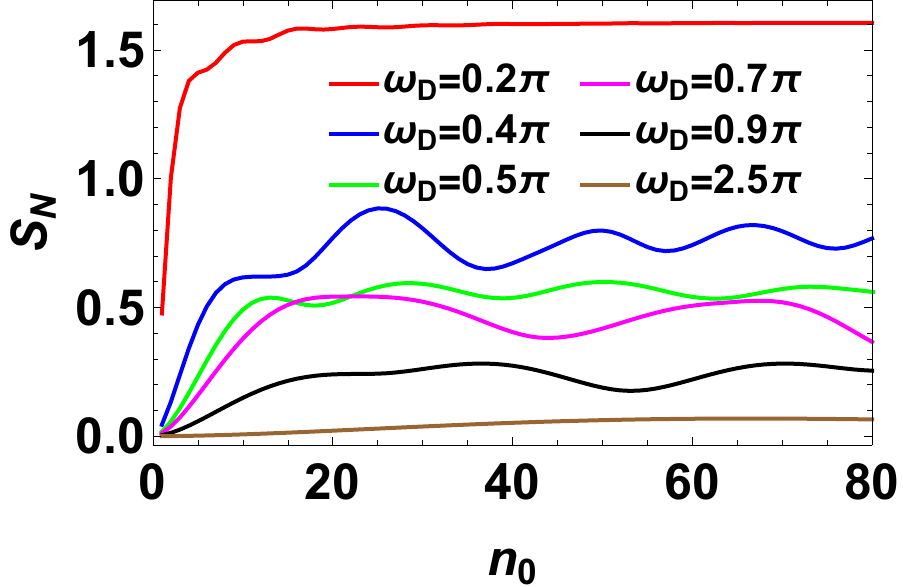}}
\caption{Left Panel: Plot of the temporal auto-correlation function
for different frequencies within the range $0.1 \le \omega_D/(\pi
\mathcal{J}_0)\le 2.5$ (shown by the legend on the right panel) as a
function of number of drive cycles $n_0$. Right Panel: Similar plot
of the number entropy $S_N$ as a function of $n_0$. All other
parameters are same as in Fig.\ \ref{fig1}. See text for details. }
\label{difffreq}
\end{figure}

We conclude this section by reinforcing how the short-time behavior
of ${\mathcal A}$ and $S_{N}$ show differences in different regions
of the drive frequency. To this end, in the left panel of Fig.\
\ref{difffreq}, we plot $\mathcal{A}$ for $L=22$ and several
representative $\omega_D$. The plot displays the nature of the decay
of ${\mathcal A}$ in different drive frequency regimes. The slope of
the decay gradually decreases with increasing frequency. The
position of the first dip also slowly shifts towards higher $n_0$
with increasing $\omega_D$. Thus we find that for sufficiently large
$n_0$, ${\mathcal A}$ settles to a frequency dependent value. The
right panel shows a plot of $S_{N}$ for the same set of parameters
and paints a similar qualitative picture. However, in contrast to
the behavior of ${\mathcal A}$, here the change of growth is much
sharper. This can be attributed to the fact that $S_{N}$ changes
from logarithmic to sub-logarithmic growth with variation of
$\omega_D$ and grows very slowly in the non-ergodic phase which is
achieved at higher $\omega_D$. We note that growth rate is frequency
independent and we shall discuss this feature in detail in the next
section.

\subsection{Intermediate-time Dynamics}
\label{longdynamics}

In this section, we discuss the dynamics of the driven interacting
AA chain at intermediate stroboscopic time, {\it i.e.}, for $n_0
\sim 10^3$. This corresponds to a time scale which is an order of
magnitude higher than that of the last section. We shall mainly
concentrate on $S_{vN}$, which shows scaling laws in this time
regime and also briefly discuss the behavior of $S_N$. Other
quantities, such as ${\mathcal A}$, do not show any additional
features and shall not be addressed here.

For this subsection, we start from the Neel or CDW initial state
given by $|\psi_0\rangle = |1010...\rangle$. Such a choice is
motivated by the results of Ref. \onlinecite{inhomquench}, where it
has been shown that the inhomogeneities in the initial state cause
changes in the $\ln t$ growth of $S_{vN}$. In fermionic systems the
most homogeneous state is expected to be the $|000..\rangle$ or
$|111... \rangle$. However such initial states does
not show any time evolution since the Floquet Hamiltonian conserves
total particle number. Thus in the particle sector $N/2$, which
constitutes the largest fraction of the Hilbert space, the most
homogeneous product state is the CDW state.
\begin{figure}[H]
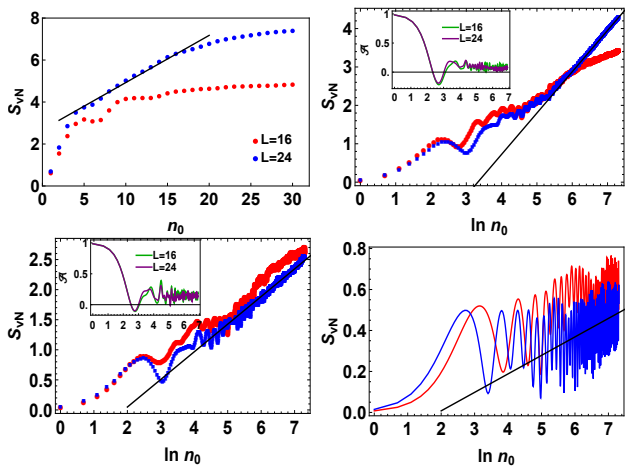

\centering{\ing[width=0.47\linewidth,height=0.35\columnwidth]{{longSV_omega0.10}.pdf}}
\centering{\ing[width=0.47\linewidth,height=0.35\columnwidth]{{longSV_omega0.45new}.pdf}}
\centering{\ing[width=0.47\linewidth,height=0.35\columnwidth]{{longSV_omega0.55new}.pdf}}
\centering{\ing[width=0.47\linewidth,height=0.35\columnwidth]{{longSV_omega1.00}.pdf}}
\caption{Top left panel: Plot of $S_{vN}$ as a function of $n_0$ for
$\omega_D/(\pi \mathcal{J}_0)= 0.1$ (ergodic regime). Top right
panel: Plot of $S_{vN}$ as a function of $\ln n_0$ for
$\omega_D/(\pi \mathcal{J}_0)= 0.4$ near the transition from the
ergodic to the multifractal regime. Bottom left panel: Same as top
right panel but for $\omega_D/(\pi \mathcal{J}_0)= 0.55$
(multifractal regime). Bottom right panel: Same as top right panel
but for $\omega_D/(\pi \mathcal{J}_0)=1.0$ (MBL phase). All other
parameters are same as in Fig.\ \ref{fig1}. See text for details.}
\label{longVN}
\end{figure}
Starting from $|\psi_0\rangle$, Fig.\ \ref{longVN} shows the growth
of $S_{vN}$ with $n_0$ for two different system sizes $L=16$ and
$24$. As seen from the top left panel for $\hbar \omega_D/(\pi
\mathcal{J}_0)= 0.1$, $S_{vN}$, in the ergodic phase, shows the
expected initial linear growth followed by saturation to a $L$
dependent value. In contrast in the MBL phase, as shown in the
bottom right panel for $\hbar \omega_D/(\pi \mathcal{J}_0)= 1.0$, we
find a $\ln t$ growth of the entanglement, albeit with a very small
slope. We note that in the high frequency regime each cycle
represents a much smaller time step; in addition, the higher
frequency also causes inherent dynamical localization \cite{dynloc3}
which stretches the time the system requires to reach the steady
state than that expected from an equilibrium MBL setup. This shows
up as large oscillations in the plot. We also find that $L=16$ shows
slightly faster growth in late times than $L=24$ for higher
frequencies. This can be attributed to local effects of the Aubr\'e
Andre potential which become prominent due to localization(both
dynamical and many-body) as frequency is increased.

For the multifractal region as seen from the top Right and the
bottom left panels of Fig.\ \ref{longVN}($\hbar \omega_D/(\pi
\mathcal{J}_0)= 0.45$ and $\hbar \omega_D/(\pi \mathcal{J}_0)= 0.55$
respectively), $S_{vN}$ is still found to follow a $\ln n_0$ growth.
We note that the $\ln n_0$ growth of $S_{vN}$ at these frequencies
shows up approximately at times after $\mathcal{A}$ has decayed
towards its long-time value. In this regime, ${\mathcal A}$
oscillates around a non-zero frequency dependent steady-state value.
This signifies presence of two different timescales in the problem
and shows that the logarithmic growth of $S_{vN}$ need not be a
feature just of MBL states for which the auto-correlation does not
decay. Instead, it can also be feature of systems which are
intermediate between ergodic and MBL. This points to a behavior
$S_{vN} \sim a \ln n_0 +b$ with $a$ and $b$ dependent on $\omega_D$.


\begin{figure}[H]
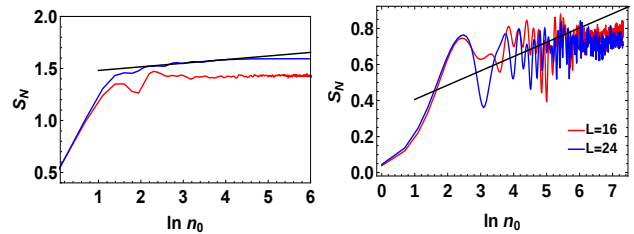

\centering
{\ing[width=0.47\linewidth,height=0.35\columnwidth]{{longSN_omega0.10}.pdf}}
\centering
{\ing[width=0.47\linewidth,height=0.35\columnwidth]{{lnSN_omega0.55}.pdf}}
\caption{Left Panel: Plot of $S_{N}$ as a function of $n_0$ for
$\hbar \omega_D/(\pi \mathcal{J}_0)= 0.1$ (ergodic regime) showing
$\ln t$ behavior. Right Panel: Plot of $S_{N}$ as a function of $
\ln n_0$ for $\omega_D/(\pi \mathcal{J}_0)= 0.55$ (multifractal
regime) showing that the number entropy grows slower than $\ln t$ in
this regime. All other parameters are same as in Fig.\ \ref{fig1}.
See text for details.} \label{longSN}
\end{figure}

Next, in Fig. \ref{longSN}, we address the behavior of $S_N$ as a
function of $n_0$. The left panel of Fig.\ \ref{longSN}, for $\hbar
\omega_D/(\pi \mathcal{J}_0)= 0.1$ (ergodic regime), we find a
logarithmic increase in $S_N$ in the same timescales where $S_{vN}$
increases linearly with $n_0$(Fig. \ref{longVN}). We find that $S_N
= 1.46+0.03 \ln n_0$ provides an accurate description of the
behavior of $S_N$ in the ergodic regime. It is to be noted that the
initial sharp rise in both $S_{vN}$ and $S_N$ is due to local
effects and is not the long-time behavior we intend to study. In
this long time regime, the growth is expected to be $ \sim \ln \ln t$ as
discussed in some recent MBL studies \cite{numberentrp2}. However,
with the numerically accessible system sizes that we have, while we
can confirm that the entanglement growth appears to be
sub-logarithmic (the black line in the right panel of Fig.\
\ref{longSN} shows a logarithmic fit), much larger system sizes and
time scales are required to determine the exact form of the growth.

\begin{figure}[H]
\centering
{\ing[width=0.47\linewidth,height=0.35\columnwidth]{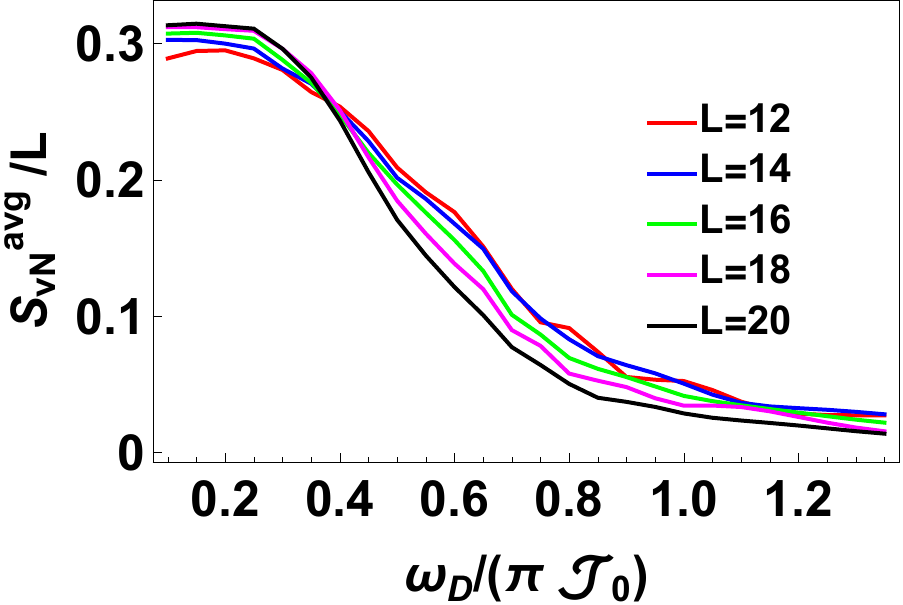}}
\centering
{\ing[width=0.47\linewidth,height=0.35\columnwidth]{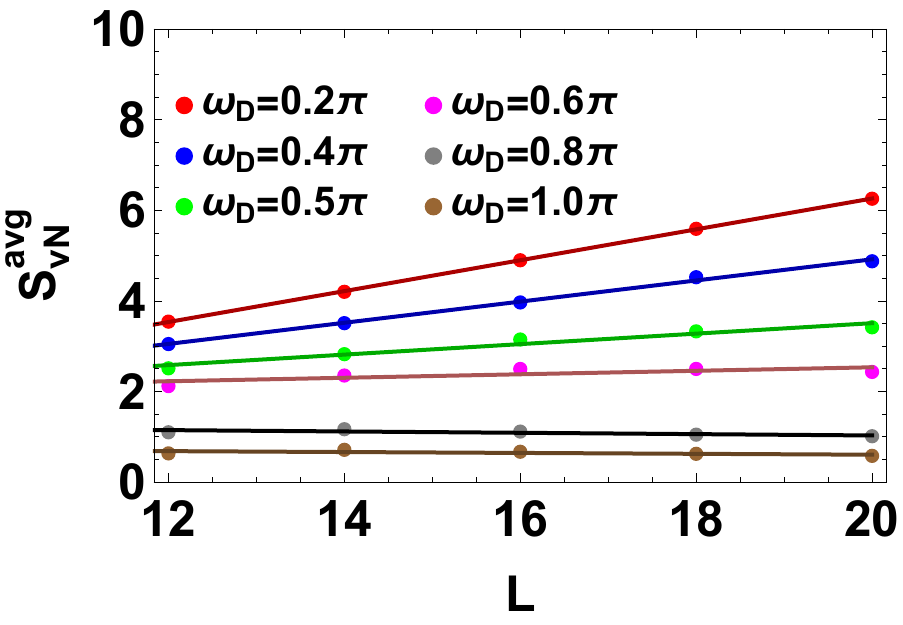}}
\caption{Left Panel: Plot of $S_{vN}/L $ averaged over ${\rm
Int}[9000/T]-{\rm Int}[10000/T]$ cycles as a function of
$\omega_D/(\pi \mathcal{J}_0)$ for different $L$. The plots indicate
crossing at $\omega_D/(\pi \mathcal{J}_0) \sim 0.45$ for different
$L$ which may signify a transition from the ergodic to multifractal
regime. Right Panel: Plot of $S_{vN}/L$ averaged over ${\rm
Int}[9000/T]-{\rm Int}[10000/T]$ cycles as a function of $L$ for
representative frequencies. All other parameters are same as in
Fig.\ \ref{fig1}. See text for details.} \label{longa}
\end{figure}

As seen from Fig.\ \ref{longVN} the evolution of entropy can be fit
to $S_{vN}(n_0) \sim a \ln n_0 + b$ where $a$ and $b$ are frequency
dependent constants. We found that $a$ decreases as the frequency
 $\omega_D$ is increased for $0.45 \le \hbar
\omega_D/(\pi \mathcal{J}_0) \le 0.8$ (in the multifractal regime).
$a$ decreases sharply after the transition (around $\hbar \omega_D/(\pi {\mathcal
J}_0)=0.43$) from ergodic to the multifractal regime. For  $\hbar
\omega_D/(\pi {\mathcal J}_0) >0.7$ $a$ approaches zero and becomes
almost independent of $\omega_D$ signifying the onset of MBL region.

Finally, we study the plot of $S^{\rm av}$, which is average of
$S_{vN}/L $ over ${\rm Int}[9000/T] \le n_0 \le {\rm Int}[10000/T]$
cycles (where ${\rm Int}[x]$ denotes nearest integer to $x$), as a
function of frequency. The corresponding plot is shown in the left
panel of Fig.\ \ref{longa}. Here, instead of looking at equal number
cycles $n_0$, we study the behavior of the quantities averaged over
equal span of stroboscopic time $n_0 T$. This is done in the regime
where $n_0 T$ is large compared to other time scales in the model.
As seen from the plot, we find a crossing around $\hbar
\omega_D/(\pi \mathcal{J}_0) \sim 0.4$. This is indicative of the
transition from the ergodic to the multifractal regime as also seen
earlier from the behavior of the Shannon entropy. The presence of
such a crossing may be understood as follows. In this MBL regime,
the system takes an exponentially long time to reach the steady
state where the average $S$ ($S^{\rm av}$) obeys a volume law. Thus,
for a fixed time, $S_{\rm av}$ decreases with system size since it
stays closer to its steady state value for smaller $L$. In contrast,
for the ergodic regime $S^{\rm av}$ reaches its steady state value
at relatively short times. Hence $S^{\rm av}$ yields the steady
state value which increases with $L$ in this regime. The fact that
one finds a crossing between $S^{\rm av}$ for different $L$ is
indicative of a length-scale independent transition point between
the ergodic and the multifractal regimes. Finally, we note that
$S^{\rm av} \sim a_0 L$ in the steady state which indicates that it
follows volume law in this regime. However, $a_0$ depends on the
drive frequency and approaches zero as we enter the MBL regime. This
is indicative of the large time scale required to approach the
steady state as discussed earlier.

\subsection{Steady state}
\label{steady}

In this section we study the steady state properties of the system
directly from the eigenfunctions of $H_F$. While in MBL regime it is
extraordinarily difficult to experimentally reach this state due to
the enormous timescales, it is still an important aspect to look at
as features embedded in $H_F$ show up most prominently in this
regime. In what follows, we shall study fermionic transport,
auto-correlation function and the number entropy in the steady
state.

\subsubsection{Transport}
\label{density}

To study transport in the system, we start from a domain wall
initial state defined in the fermion number basis by,
\begin{eqnarray}
|\psi_{\rm init}\rangle = |n_{1}=1, ... n_{L/2}=1, n_{L/2+1}=0, ...
n_{L}=0\rangle  \label{initstate}
\end{eqnarray}
where the system size $L$ was considered to be an even integer
(chain with even number of sites) and $n_j =\langle \hat n_j
\rangle$ denotes fermion occupation number on the $j^{\rm th}$ site.
The wavefunction after $n$ drive cycles is then given by
\begin{eqnarray} |\psi'\rangle &=& U(nT,0) |\psi_{\rm init}\rangle=
\sum_m c_m^{\rm init} e^{-i n\epsilon_m^F T/\hbar} |\psi_m\rangle
\label{finalstate}
\end{eqnarray}
where $|\psi_m\rangle$ denotes Floquet eigenstates with $L/2$
fermions and $c_m^{\rm init}= \langle \psi_m|\psi_{\rm
init}\rangle$. Using this state we study the following quantities in
order to further establish the MBL transition.
\begin{eqnarray}
N_{0j}(T) &=& \langle 2(\hat n_j-1/2)\rangle \nonumber \\
\quad N_{\rm av} (T) &=& \frac{4}{L} \sum_{j=1..L}  \langle \hat
(\hat n_j-1/2)\rangle ^2 \label{nnsq}
\end{eqnarray}
where the average is taken with respect to the steady state reached
under a Floquet drive starting from $|\psi_{\rm init}\rangle$. In
terms of the Floquet eigenfunctions $|\psi_m\rangle$ and the overlap
coefficients $c_m^{\rm init}$ (Eq.\ \ref{finalstate}) these can be
expressed as
\begin{eqnarray}
N_{0j}(T) &=& \sum_m |c_{m}^{\rm init}|^2  \langle \psi_m| 2(\hat
n_j
-1/2)|\psi_m\rangle \label{ssopexp} \\
 N_{\rm av}(T) &=& \frac{4}{L} \sum_{j=1..L} (\sum_m |c_m^{\rm init}|^2  \langle \psi_m| \hat
(\hat n_j-1/2) |\psi_m\rangle)^2 \nonumber
\end{eqnarray}
We note that for the initial state $ | \langle \psi_{\rm init}| \hat
2(\hat n_j -1/2) |\psi_{\rm init}\rangle |^2 =-1$ for j $<$ L/2 and $| \langle
\psi_{\rm init}| \hat 2(\hat n_j -1/2) |\psi_{\rm init}\rangle |^2
=1$ for j $>$ L/2, while for free fermions, the ground state with $\mathcal{J}_0
\gg V_0$, $\langle 2(\hat n_j -1/2) \rangle=0$. Thus $N_{\rm av}(T)$
provides a measure of degree of delocalization of the driven chain.
A similar reasoning shows that $N_{0j} \to 0$ for all sites in the
delocalized regime and $N_{0j} = 1[-1]$ for $j<[>]L/2$ in the
localized regime; in contrast, in the presence of a mobility edge,
$N_{0j}$ takes values between $0$ and $1$ at different sites.
\begin{figure}
\rotatebox{0}{\includegraphics*[width= 0.48 \linewidth]{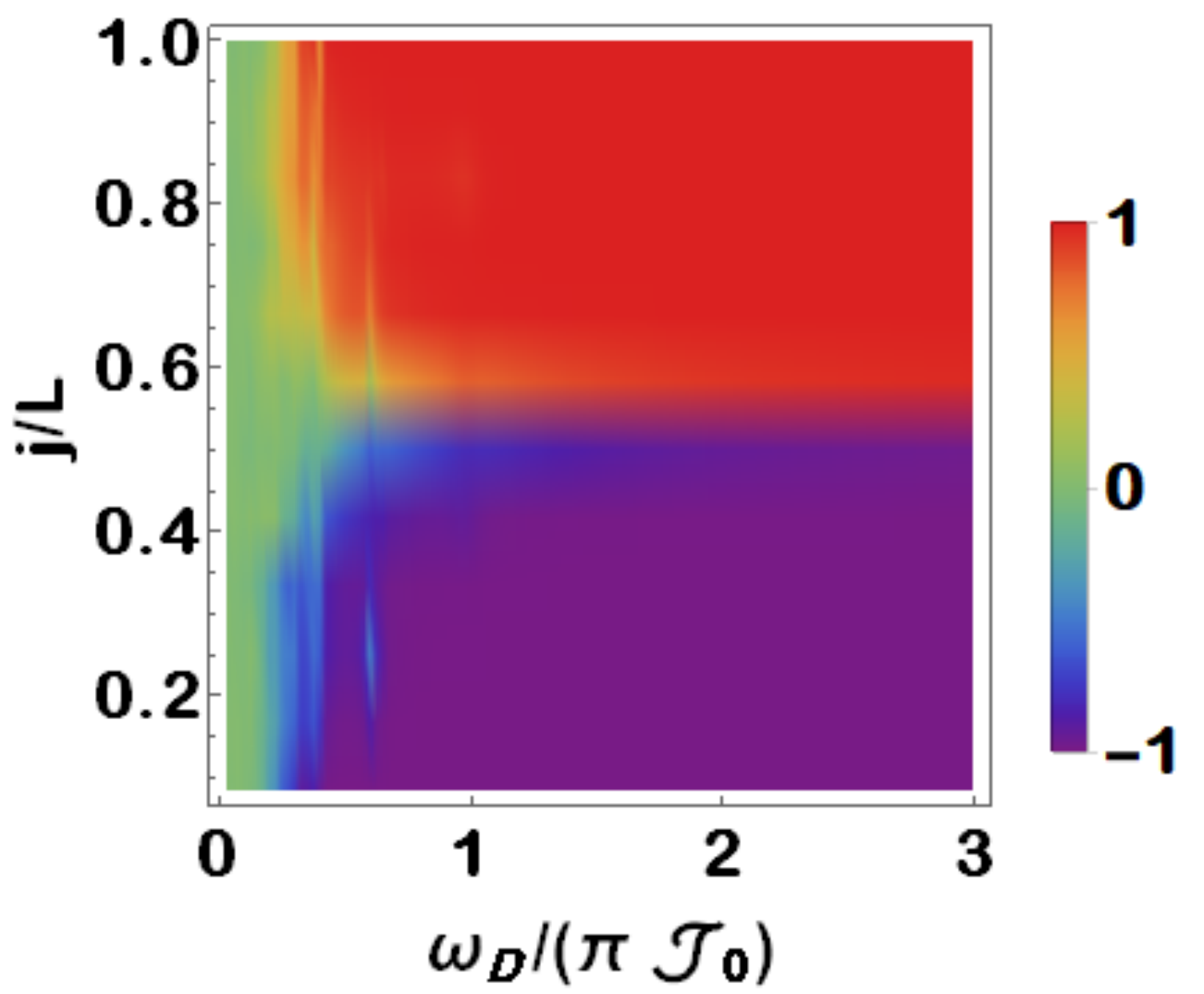}}
\rotatebox{0}{\includegraphics*[width= 0.48 \linewidth]{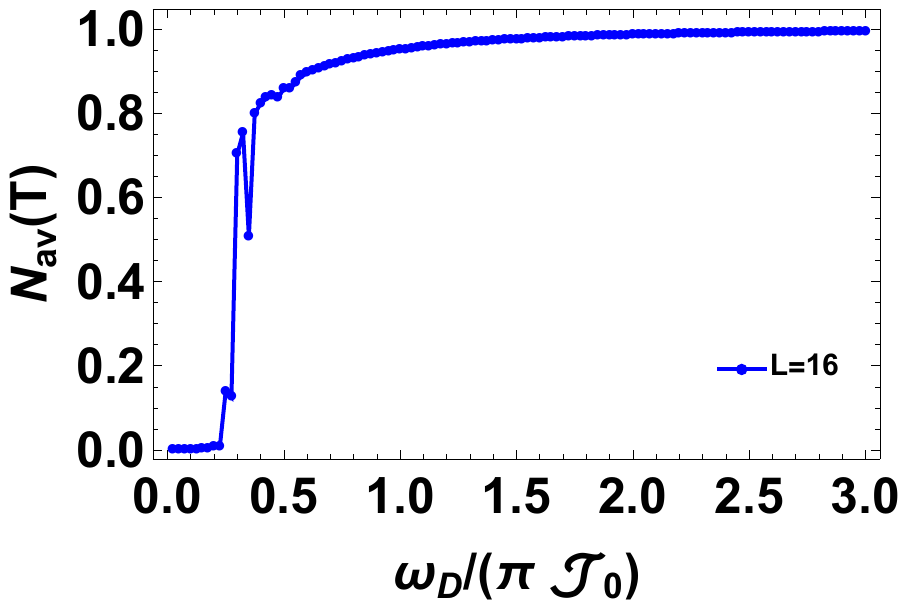}}
\caption{Left: Plot of $N_0$ as a function of $j/L$ and
$\omega_D/(\pi\mathcal{J}_0)$ showing fermion density profile at all
sites of the chain in the steady state as a function of
$\omega_D/(\pi\mathcal{J}_0)$. Right: Plot of $N_{\rm av}(T)$ as a
function of $\omega_D/(\pi\mathcal{J}_0)$ in the steady state
showing $0 \le N_{\rm av}(T) \le 1$ for $0.3 \le \omega_D/(\pi
\mathcal{J}_0) \le 1.5$. All other parameters are same as in Fig.\
\ref{fig1}. See text for details.} \label{density}
\end{figure}

A plot of $N_0$ as a function of $j/L$ and $\omega_D$, shown in the
left panel Fig.\ \ref{density}, indicates that the transition from
the ergodic to the multifractal  and MBL regions leaves it signature
in fermion transport. We find that the steady state value of $N_0$
in the MBL (high frequency) regime is $\sim \pm 1$ for the left and
right halves of the chain respectively. This indicates that the
steady state is close to the initial state as expected in the MBL
regime. In contrast, $N_0=0$ for all $j/L$ in the ergodic (low
frequency) regime which indicates that the system has reached the
infinite temperature steady state as expected for a driven ergodic
many-body system. In between, the system displays a range values of
$N_0$ for different $j$ which indicates the intermediate behavior in
the multifractal regime. This is also reflected in the plot of
$N_{\rm av}$ which shows a kink in the multifractal regime and
indicates a transition between the localized and delocalized
regimes. We note here that the steady state localization here
happens due to both MBL nature of the Floquet eigenstates and
dynamical localization due to the drive; thus fermion transport may
not solely reflect MBL properties in the high frequency regime.

\subsubsection{Auto-correlation function}

We define the steady state auto correlation function at site $j$ as,
\begin{eqnarray}
\mathcal{A}_s^j = ( 2 n_j^s -1 )( 2 n_j^0  -1 )
\end{eqnarray}
where ,
\begin{equation*}
n_j^s = \sum_m |c_{m}^{\rm init}|^2  \langle \psi_m| \hat
n_j|\psi_m\rangle
\end{equation*}
the steady state value of $\langle \hat n_j\rangle$ and $n_j^0 =
\langle \psi_{\rm init}|\hat n_j|\psi_{\rm init}\rangle$ is the
initial value. We average this single site operator over different
sites to compute $\mathcal{A}_s^{\rm
av}=\frac{1}{L}\sum_j\mathcal{A}_s^j$.
\begin{figure}
\rotatebox{0}{\includegraphics*[width= 0.48
\linewidth]{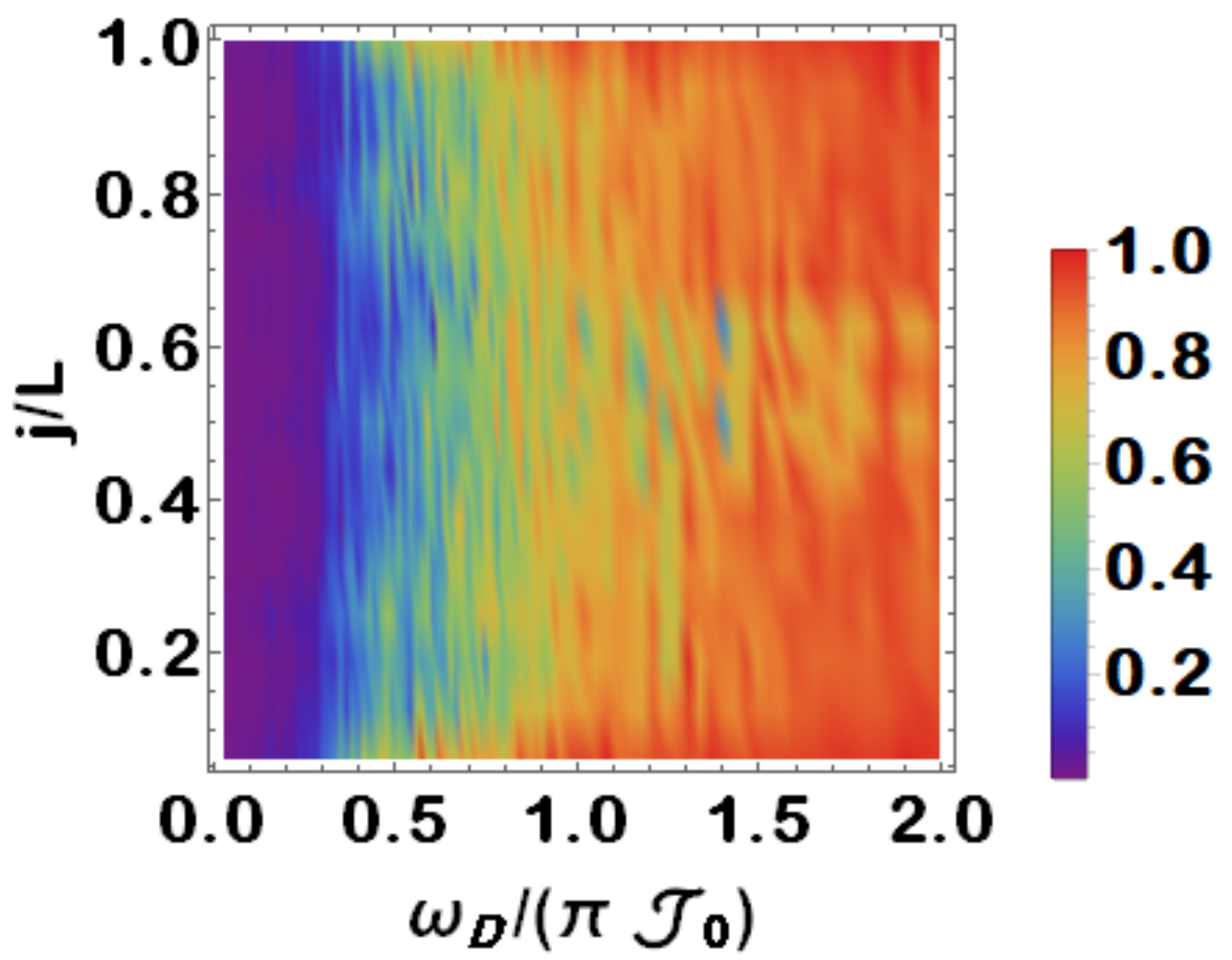}}
\rotatebox{0}{\includegraphics*[width= 0.48
\linewidth]{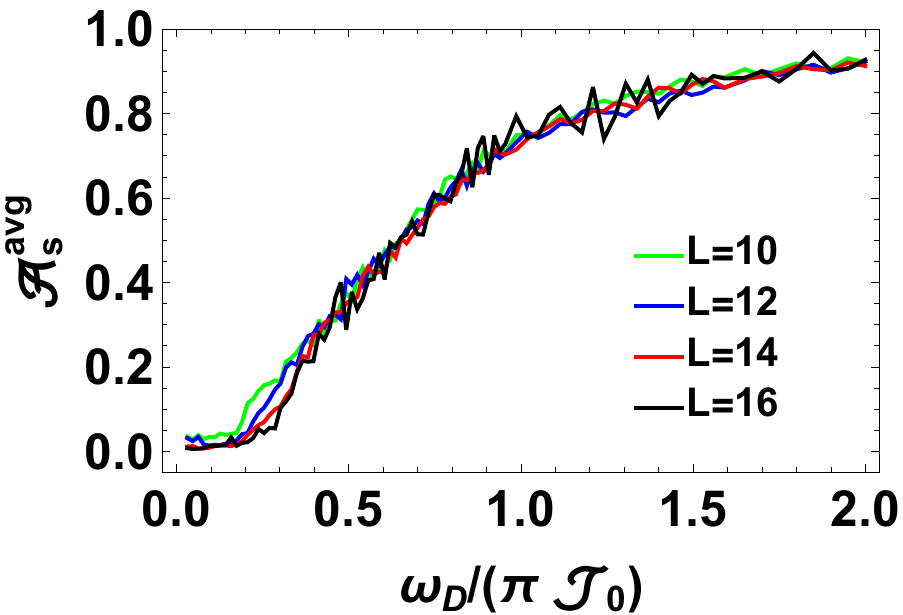}} \caption{Left: Plot of $\mathcal{A}_s$
as a function of $j/L$ and $\omega_D/(\pi\mathcal{J}_0)$ showing the
auto-correlation at all sites of the chain in the steady state as a
function of $\omega_D/(\pi\mathcal{J}_0)$. Right: Plot of
$\mathcal{A}_s^{\rm avg}$ as a function of
$\omega_D/(\pi\mathcal{J}_0)$. All other parameters are same as in
Fig.\ \ref{fig1}. See text for details.} \label{steadyauto}
\end{figure}

The left panel of Fig.\ \ref{steadyauto} shows the plot of steady
state value of the auto-correlation function as a function of $j$
and $\omega_D$. This value is computed by averaging over $N_0=50$
product initial states in the basis of $H$ for $L=10-14$ and $N_0
=10$ such states for $L=16$. Below $\hbar \omega_D/(\pi
\mathcal{J}_0) \sim 0.25$, the value of autocorrelator remains zero
indicating $n_j^s=1/2$ for all $j$ in the steady state. In contrast
at high drive frequencies, one expects $n_j^s \simeq n_j^0$ leading
to ${\mathcal A}^{\rm av} \simeq 1$. In between $ 0.25 < \hbar
\omega_D/(\pi \mathcal{J}_0) < 1 $, the behavior is intermediate to
that of delocalized or MBL phase. Thus $0 < {\mathcal A}_s < 1$
indicates that the system is in a multifractal phase. As the
frequency is increased beyond that, $\mathcal{A}_s \to 1$ indicating
the onset of localization. As before, we point out that this
localization receives contribution from both the MBL nature of the
Floquet eigenstates and the dynamical localization due to the drive.

\subsubsection{Number entropy}

In this subsection we shall show the steady state behavior of $S_N$
denoted by $S_N^{s}$. We divide the system into two subsystems $A$
and $B$ and we integrate over subsystem $B$. To compute the number
entropy, we first denote the states in the fermion number basis as
$|\chi_k\rangle$. Since these are eigenstates of the number
operator, for each of them, one can compute the total number of
fermions $n$ in subsystem A: $n_{A k}= \sum_{j \in A} \langle
\chi_k|\hat n_j|\chi_k\rangle$. Using the notation of Eq.
\ref{finalstate}, we can write in the steady state
\begin{eqnarray}
|d_k^{s}|^2 &=& |\langle\chi_k|\psi_m\rangle|^2 = \sum_m |c_m^{\rm init}|^2
|\langle \chi_k|\psi_m\rangle|^2 \label{dsdef}
\end{eqnarray}
Using Eq.\ \ref{dsdef}, we can then obtain
\begin{eqnarray}
p_{s}(n) &=& \sum_{k=k^{\prime}} |d^s_k|^2 \label{psdef} \\
S_N^{s}&=&-\sum_n p_{s}(n) \ln p_{s}(n), \,\, \langle S_N^s\rangle =
\frac{1}{N}\sum_i^N S_N^s(i) \nonumber
\end{eqnarray}
where $k^{\prime}$ denotes all states with $n_{Ak}=n$. Here
$\langle S_N^s\rangle$ denotes averaged number entropy where the
average is taken over $N$ product states. For an
initial Neel state, we denote the entropy by $\mathcal{S}_N^s$ since
no averaging is involved.
\begin{figure}
\centering
{\ing[width=0.47\linewidth,height=0.35\columnwidth]{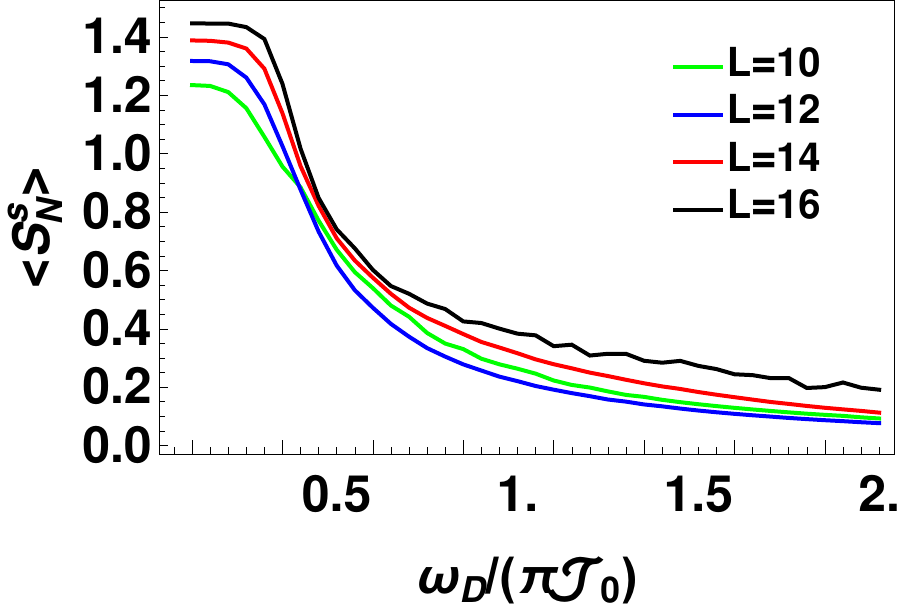}}
\centering
{\ing[width=0.47\linewidth,height=0.35\columnwidth]{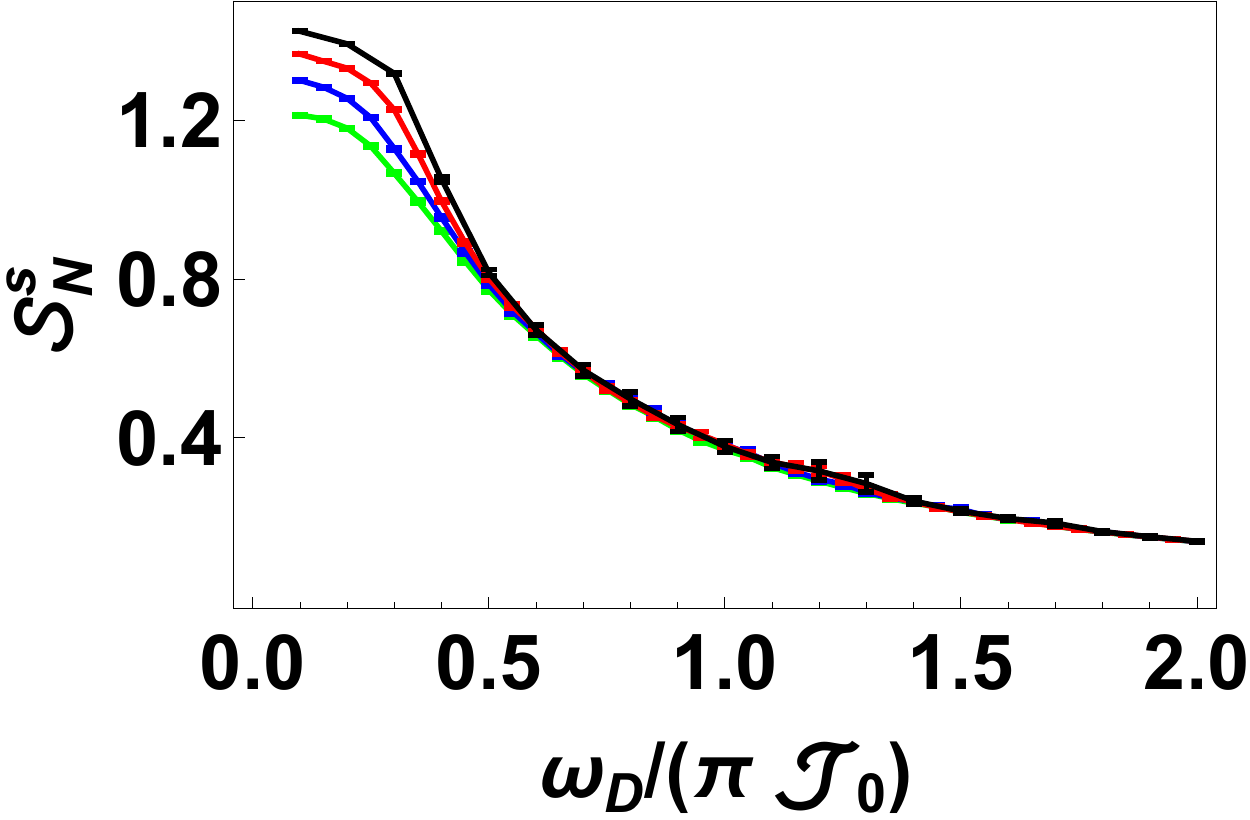}}
\caption{Left: Plot of $\langle S_N^s \rangle$ as a function of
$\omega_D/(\pi\mathcal{J}_0)$ showing the decrease in number entropy
with increasing frequency For sizes $L=10,12,14$ we have averaged
over all possible product initial states, for $L=16$ we have
averaged over several random product states such that the error bar
is smaller than line width. Right: Plot of $S_N$ vs
$\omega_D/(\pi\mathcal{J}_0)$ for $|\psi_{\rm init}\rangle$ being
the Neel state.  All other parameters are same as in Fig.\
\ref{fig1}. See text for details.} \label{steadySN}
\end{figure}

Fig. \ref{steadySN} shows the steady state behavior of $S_N$ for
different drive frequencies. The left panel shows the average
behavior when we take $\ell$ product states: $|\psi_{\rm
init}\rangle=|\chi_{\ell}\rangle$. We compute $\langle S_N^s
\rangle$ by averaging over these states. In the low frequency regime
where ergodicity is expected, $\langle S_N^s\rangle$ is large and
monotonically increases with $L$. However with increasing frequency
as the system becomes non-ergodic, it decreases and there is no
clear monotonicity with $L$. This is similar to the behavior seen in
Ref.\ \onlinecite{numberentrp2}. A similar behavior is also seen on
the right panel where $|\psi_{\rm init}\rangle$ is taken to be the
Neel state. For these plots, we have chosen the AA potential to be
$V_j= V_0 \cos(2 \pi \eta j +\phi)$ and have averaged over $\phi$ to
smoothen out possible local fluctuations. These local fluctuations
tend to arise in $S_N$ at high frequencies as the steady state
values heavily depend on the local potentials near the half-chain
cut. To prevent this from affecting our overall result we perform
the averaging in this scenario. From the plot, we find that at large
drive frequencies ({\it i.e}., in the localized regime), the steady
state curves for different $L$ almost overlap. In contrast, in the
ergodic regime there is a clear increase of $S_N$ with system size.
The plot confirms that the system is localized at high frequencies
but not at intermediate frequencies.

\section{Floquet Perturbation Theory}
\label{fptsec}

In this section, we aim to obtain a semi-analytic, albeit
perturbative, understanding of several features of the driven
interacting AA model found via exact numerics using a Floquet
perturbation theory which is known to provide accurate results  at
intermediate frequencies provided that the term in $H(t)$ with
largest amplitude is treated exactly\cite{dynloc3,rev9}. In the
present case, since ${\mathcal J}_0 \gg V_0, V_{\rm int}$, one needs
to treat the drive term exactly. Thus one obtains
\begin{eqnarray}
U_0(t,0) &=&  e^{i {\mathcal J}_0 t \sum_k \epsilon_k \hat n_k,
/\hbar} \quad t \le T/2
\nonumber\\
&=& e^{i{\mathcal J}_0(T-t) \sum_k \epsilon_k \hat n_k /\hbar} \quad
t\ge T/2  \label{zeroord}
\end{eqnarray}
where $U_0$ is the exact evolution operator corresponding to $H=
H_0= {\mathcal J}_0 \sum_k \epsilon_k \hat n_k$ and $\epsilon_k=
-2\cos k$ for the fermion chain with nearest-neighbor hopping. From
Eq.\ \ref{zeroord}, we find that $U_0(T,0)= I$ indicating
$H_F^{(0)}=0$.

Next, we compute the first order Floquet Hamiltonian $H_F^{(1)}$. To
this end, we note that within first order perturbation theory, the
correction to the evolution operator is given by \cite{rev9}
\begin{eqnarray}
U_1(T,0) &=&  \frac{-i}{\hbar} \int_0^T dt U_0^{\dagger}(t,0) (H_1 +
H_A) U_0 (t,0)  \nonumber\\
H_{F}^{(1)} &=& \frac{i}{T} U_1(T,0) \label{firstord}
\end{eqnarray}
The computation of $H_{F}^{(1)}$ can be done in a straightforward manner
following the method discussed in Refs.\ \onlinecite{dynloc3,rev9}.
The matrix elements of $H_{F}^{(1)}$ between Fock states in momentum
space, denoted by $|n_k\rangle = |n^{(1)}_{k_1} .....
n^{(N)}_{k_N}\rangle$, is given by
\begin{widetext}
\begin{eqnarray}
\bra{n_{i}} H_F^{(1)}\ket{n_{f}}&=& V_0 \sum_{q,k_1} \frac{2\hbar
(1-e^{ i T \mathcal{J}_0 \alpha_1(k_1,q)/\hbar})}{T
\mathcal{J}_0 \alpha_1(k_1,q)} f(q)\delta_{n^{i}_{k_1}, n^{f}_{k_1+q}-1} \nonumber\\
&&+V_{\rm int} \sum_{k_1,k_2,q}\frac{2 \hbar(1-e^{ i T \mathcal{J}_0
\alpha_2(k_1,k_2,q)/\hbar})}{T  \mathcal{J}_0 \alpha_2(k_1,k_2,q)}
e^{-i q}
\delta_{n_{i}^{k_1},n_f^{k_1}-1}\delta_{n_{i}^{k_2},n_f^{k_2}-1}
\delta_{n_{f}^{k_2-q},n_i^{k_2-q}+1}\delta_{n_{f}^{k_1+q},n_i^{k_1+q}+1}
\nonumber\\
\alpha_1(k_1,q) &=& \cos k_1-\cos (k_1+q), \quad
\alpha_2(k_1,k_2,q)=\cos k_1+\cos k_2-\cos (k_1+q)-\cos(k_2-q)
\label{FPTmom}
\end{eqnarray}
\end{widetext}
where $f(q)= \sum_j \exp[-i j q] \cos(2 \pi \eta j)$. We note that
for $T \to 0$, the Floquet Hamiltonian reduces to that obtained from
the first order Magnus expansion $H_F^{(1)}(T \to 0) \simeq H_F^{\rm
magnus} = H_A + H_1$. However at intermediate frequencies, the
frequency dependence of $H_F^{(1)}$ is much more complicated.
Moreover, a similar calculation shows that $U_2(T,0)= U_1(T,0)^2/2$;
thus $H_F^{(2)}=0$ and $H_F^{(1)}$ represents the sole contribution
to $H_F$ to ${\rm O}(T^2)$. These features allow one to expect that
it shall provide at least qualitatively accurate description of the
dynamics of the system at intermediate drive frequencies.

Next, we use Eq. \ref{FPTmom} to numerically compute matrix elements
of $H_F^{(1)}$ between Fock states in the position basis. A
numerical diagonalization of the matrix thus obtained yields the
eigenvalues and eigenvectors (in the real space Fock basis) for
comparison with our exact results.
\begin{figure}
\centering{\ing[width=0.42\linewidth,height=0.35\columnwidth]{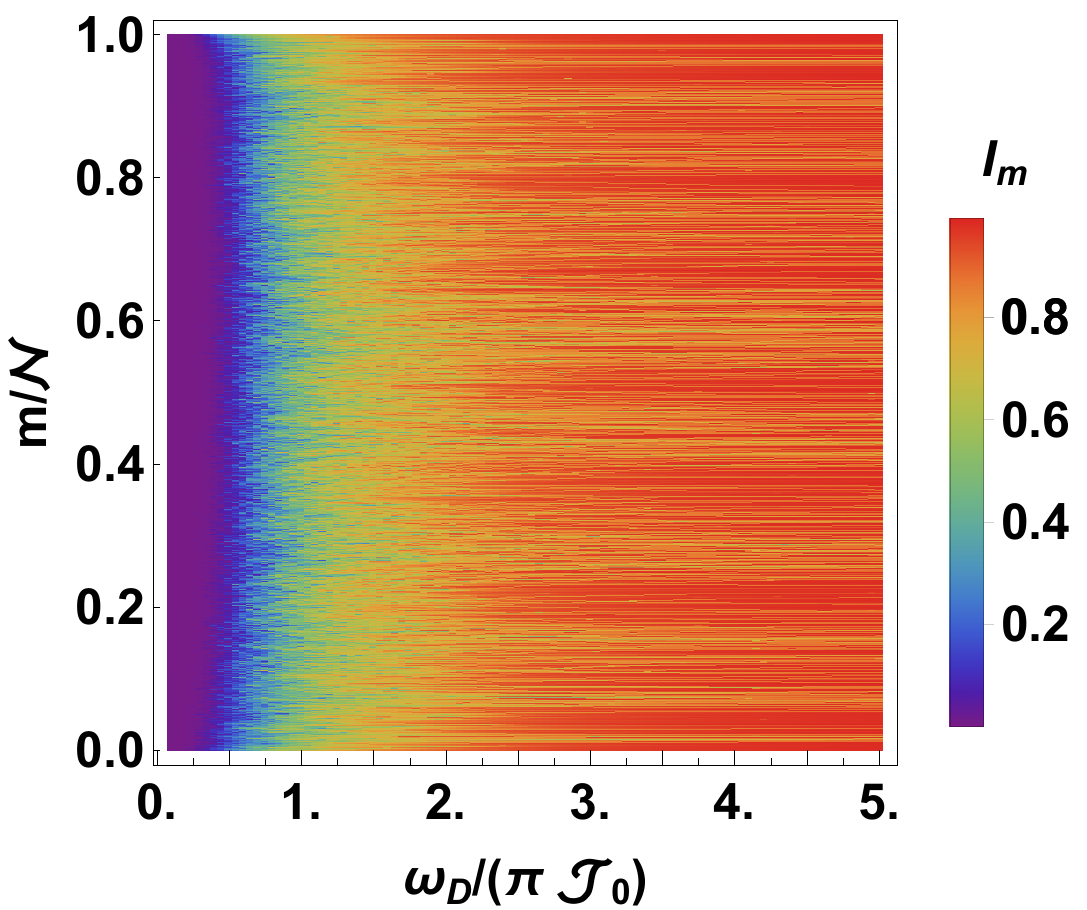}}
\centering{\ing[width=0.42\linewidth,height=0.35\columnwidth]{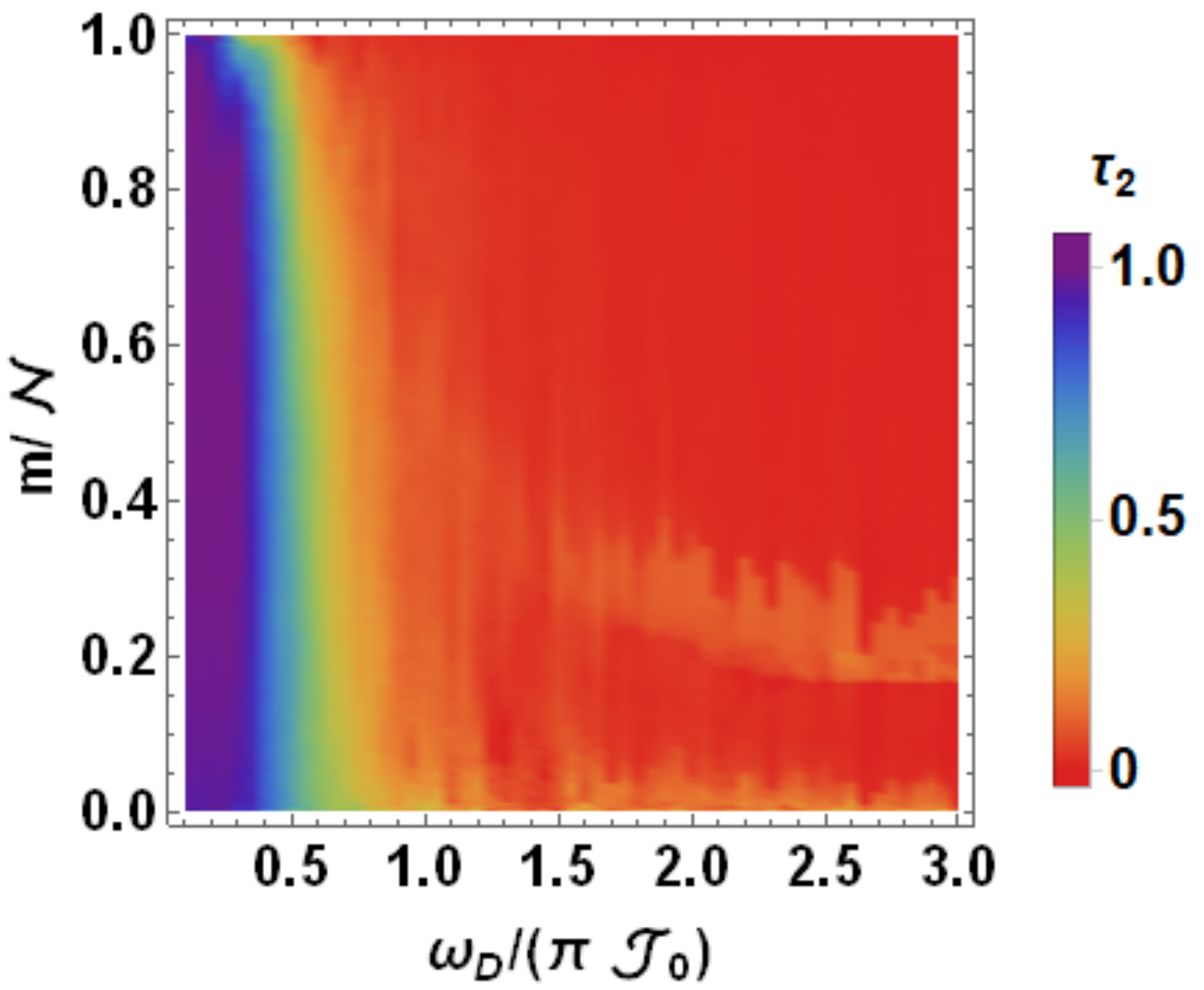}}
\centering{\ing[width=0.40\linewidth,height=0.28\columnwidth]{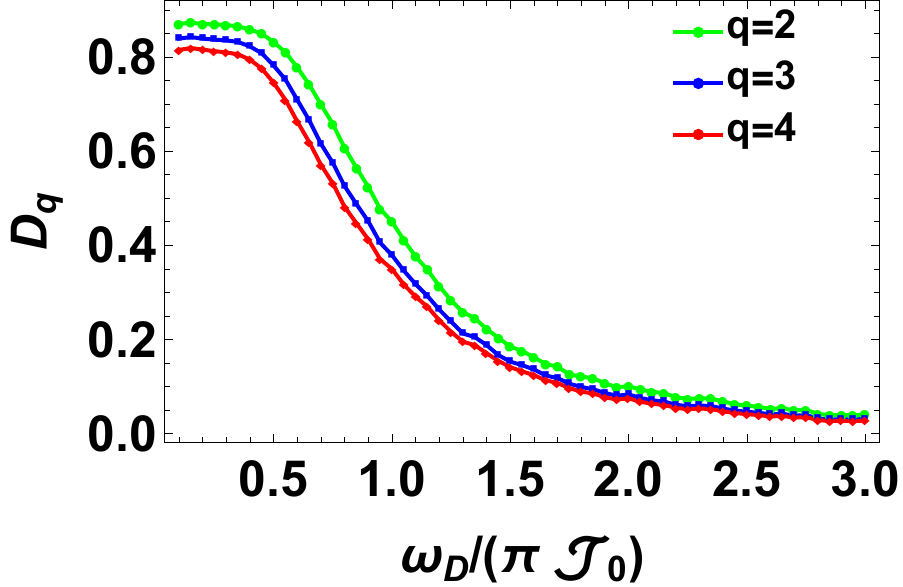}}
\centering{\ing[width=0.40\linewidth,height=0.28\columnwidth]{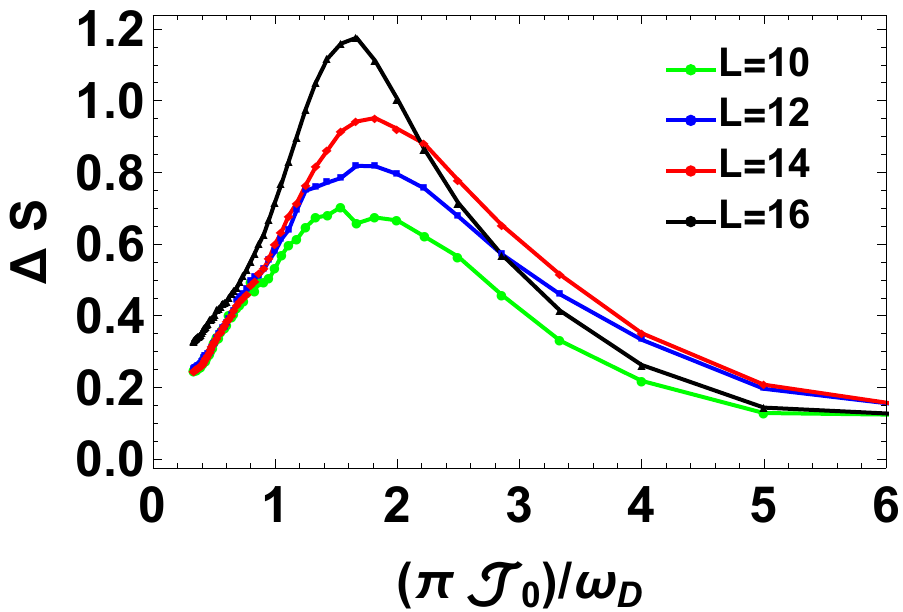}}
\caption{Top Left: Plot of $I_m$ as a function of the normalized
many-body eigenfunction index $m/\mathcal{N}$ and $\omega_D/(\pi
\mathcal{J}_0)$ showing the localized/delocalized nature of the
Floquet eigenstates $|\psi_m\rangle$ for $L=14$. Top Right: Plot of
$\tau_2$ as a function of $m/\mathcal{N}$ (after sorting in
increasing order of $I_m$) and $\omega_D/(\pi \mathcal{J}_0)$
showing the presence of delocalized states for $ \omega_D/(\pi
\mathcal{J}_0)\le 0.45$, multifractal states for $0.45  \le
\omega_D/(\pi \mathcal{J}_0)\le 1.5$ and localized states for $
\omega_D/(\pi \mathcal{J}_0) > 1.5$  . The system sizes used for
extracting $\tau_2$ are $L=10, \cdots, 16$ in steps of $2$. Bottom
Left : Plot of $D_q$ as a function of $\omega_D/(\pi \mathcal{J}_0)$
for $m/\mathcal{N}=0.5$. Bottom Right: Plot of $ \Delta S$ as a
function of $(\pi\mathcal{J}_0)/\omega_D$. All these plots has been
done using FPT and parameters are same as in Fig.\ \ref{fig1}. See
text for details.  } \label{fpt}
\end{figure}

The results obtained from the above-mentioned procedure is depicted
in Fig.\ \ref{fpt}. From Fig.\ \ref{fpt} we find that the results
obtained from FPT agrees with those from exact numerics discussed in
Sec. \ref{phase}. The top left panel of Fig.\ \ref{fpt} shows the
plot of $I_m$, obtained from eigenvectors of $H_F^{(1)}$ using Eq.\
\ref{ipr1}, as a function of $\omega_D$. The top right panel shows
the corresponding plot for $\tau_2$. We find that both the plots
show similar multifractal behavior as seen in Figs. \ref{fig1} and
\ref{fig2} within similar range of $\omega_D/(\pi \mathcal{J}_0)$.
In particular, we find that the eigenvectors of $H_F^{(1)}$ exhibits
delocalized eigenstates for $ \omega_D/(\pi \mathcal{J}_0)\le 0.45$,
multifractal eigenstates states for $0.45 \le \omega_D/(\pi
\mathcal{J}_0)\le 1.5$ and localized eigenstates for $ \omega_D/(\pi
\mathcal{J}_0) > 1.5$. The plot of $D_q$ as a function of
$\omega_D$, shown in the bottom left panel of Fig.\ \ref{fpt}, also
shows qualitatively similar behavior to that obtained from ED shown
in bottom right panel of Fig.\ \ref{fig1}; however, we note that the
change in $D_q$ signifying the transition to multifractal phase is
seen around $\hbar \omega_D/(\pi {\mathcal J}_0) \simeq 0.6$. Thus
the position of the transition is not accurately captured by
$H_F^{(1)}$. Nevertheless, $H_F^{(1)}$ does predict the transition
to the multifractal phase as seen in the bottom right panel of Fig.\
\ref{fpt}, where a plot of $\Delta S$ as a function of $\omega_D$
shows a distinct peak. The peak becomes sharper with increasing $L$
which is consistent with the result obtained from ED.
Thus we conclude that $H_F^{(1)}$, computed using FPT,
constitutes a semi-analytic Floquet Hamiltonian which shows a
transition from ergodic to multifractal regime at intermediate
frequencies.

\section{Discussion}
\label{discussion}

In this work, we have studied a driven fermionic chain with an AA
potential and nearest neighbor density-density interaction between
the fermions. Our analysis constitutes a detailed study, both
numerical and semi-analytic, of the Floquet Hamiltonian of such a
system as a function of drive frequency in the limit of large drive
amplitude.

We have shown that such a driven system supports multifractal
many-body Floquet eigenstates for a range of drive frequencies in
the intermediate drive frequency regime. We find that the
eigenstates are ergodic in the low frequency and many-body localized
in the high drive frequency regime. In between, for system sizes
accessible in our numerics, results indicate a possible transition
from ergodic to multifractal phase at $\omega_D= \omega_c \simeq
0.43 {\mathcal J}/\hbar$. Upon further increasing the drive
frequency, the eigenstates become many-body localized via a smooth
crossover. The presence of the transition from ergodic to the
multifractal phase seems likely due to two reasons. First, the plot
of entropy fluctuation $\Delta S$ as a function of $T$ shows a peak
at the transition which gets sharper with increasing system size.
Second, $b_1$ (Eq.\ \ref{enteq}) changes sign at this point which is
indicative of a transition from the ergodic phase. However, we need
that one needs finite size numerics with larger system size to
settle this issue; this has not been attempted in this work.

Our analysis indicates that several dynamic quantities that we study
can distinguish between the multifractal ergodic and MBL phases.
These include the fermion auto-correlation function and the short
time behavior of the normalized participation ratio. The former
quantity decays sharply to zero in the ergodic phase due to
spreading of the system in the Hilbert space. In the MBL phase, it
remains close to its initial value since the system retains its
initial memory. In the multifractal phase, we find an initial sharp
decay of the auto-correlation function, followed by oscillation
around a steady state value which is intermediate between its
ergodic (zero) and the MBL (unity) counterparts. The latter quantity
shows system-size independent behavior as a function of number of
drive cycles in the ergodic phase and displays a clear $L$
dependence in the MBL phase. In contrast, it shows intermediate
behavior with oscillations as a function of $n_0$ in the
multifractal phase. We note that in contrast, the entanglement
entropy and the number entropy do not distinguish between the
multifractal and the MBL eigenstates.

We have also studied steady state properties of the driven system,
starting from a domain wall initial state, by computing transport
properties, auto-correlation function and the number entropy. We
find all of these quantities reflect a change from localized to
delocalized regime as a function of drive frequency. However, the
localization seen in transport also receives contribution from
dynamical localization at high drive frequencies \cite{dynloc3}. We
also find that near the transition frequency, the distribution of
the number density of fermions in the steady state acquires a large
width; this suggests a possible signature of the multifractal regime
in fermion transport. A similar feature is seen in the steady state
value of auto-correlation function which satisfies $0 < {\mathcal
A}_s <1$ in the multifractal phase; this is in sharp contrast to its
values zero and unity in the ergodic and MBL phases respectively.
The plot of steady state number entropy also show a sharp drop at
the transition which becomes sharper with increasing $L$.

We have also obtained similar qualitative features for the driven
fermionic chain from a semi-analytic, albeit perturbative, Floquet
Hamiltonian computed using FPT. Remarkably, this perturbative
Floquet Hamiltonian reproduces multifractality of the Floquet
eigenstates and also points towards a transition from the ergodic to
the multifractal regime. Our results thus constitutes an analytic
Floquet Hamiltonian which support ergodic, multifractal and MBL
eigenstates depending on the drive frequency.

Our results could be relevant for ultracold interacting fermions in
the presence of an 1D optical lattice \cite{rev8}. The realization
of the AA potential can be done using techniques discussed in Refs.\
\onlinecite{exp2} and \onlinecite{exp22}. The drive can be
implemented by appropriate tuning of the strength of the laser used
to create the optical lattice. We suggest measurement of
density-density auto-correlation of the fermions. Our results
suggest that the short time behavior of this auto-correlation
function would be sufficient to distinguish between the ergodic, MBL
and the multifractal phases. In particular, in the intermediate
drive frequency regime, we expect the auto-correlation function to
exhibit a sharp drop followed by oscillations around a finite
non-zero value.

The fate of the multifractal phase that we obtain in the
thermodynamic limit remains an open question. The phase remains
stable within the system sizes that we could access within ED;
however, it is possible that it might either shrink for large $L$
leading to a direct ergodic-MBL quantum phase transition. An
investigation of the stability of the multifractal phase in the
thermodynamic limit is beyond the scope of the present paper. We
note however, that the finite-sized chains that we study in this
paper may possibly be experimentally realized using ultracold atoms
in optical lattices.

\begin{acknowledgements}
R.G. thanks M. \u Znidari\u c for discussions and funding from project J1-1698 Many-body transport engineering for financial support.
\end{acknowledgements}

\end{document}